\documentclass[aps,prresearch,twocolumn,groupedaddress,nofootinbib,10pt,longbibliography]{revtex4-1}
\usepackage{CJK}

\usepackage{hyperref}
\usepackage{xcolor}
\usepackage{braket}
\usepackage{slashed}

\usepackage{mathtools}
\usepackage{multirow}
\usepackage{nicefrac}
\usepackage{amssymb}
\usepackage{mathrsfs}
\usepackage{IEEEtrantools}
\usepackage{tabularx}

\definecolor{BrickRed}{cmyk}{0, .89, .94, .28}

\begin{document}

\title{Modelling charge transport in gold nanogranular films}

\author{Miquel L\'opez-Su\'arez}
\email{mlopez@dsf.unica.it}
\affiliation{Dipartimento di Fisica, Università degli Studi di Cagliari, Cittadella Universitaria, I-09042 Monserrato, Cagliari, Italy}
\author{Claudio Melis}
\affiliation{Dipartimento di Fisica, Università degli Studi di Cagliari, Cittadella Universitaria, I-09042 Monserrato, Cagliari, Italy}
\author{Luciano Colombo}
\affiliation{Dipartimento di Fisica, Università degli Studi di Cagliari, Cittadella Universitaria, I-09042 Monserrato, Cagliari, Italy}
\author{Walter Tarantino}
\affiliation{Dipartimento di Fisica, Università degli Studi di Cagliari, Cittadella Universitaria, I-09042 Monserrato, Cagliari, Italy}

\date{\today}

\begin{abstract}
Cluster-assembled metallic films show interesting electrical properties, 
both in the near-to-percolation regime, when deposited clusters 
do not form a complete layer yet, and when the film thickness 
is well above the electrical percolation threshold. 
Correctly estimating their electrical conductivity is crucial, 
but, particularly for the latter regime, 
standard theoretical tools are not quite adequate. 
We therefore developed a procedure based on an atomically 
informed mesoscopic model in which ab-initio estimates 
of electronic transport at the nanoscale are used to reconstruct 
the conductivity of nanogranular gold films generated by molecular dynamics. An equivalent resistor network 
is developed, appropriately accounting for ballistic transport. 
The method is shown to correctly capture the non-monotonic behavior of the conductivity as a function of the film thickness, namely a signature feature of nanogranular films.
\end{abstract}

\maketitle
\section{Introduction}

Cluster assembled metallic (or, simply, “nanogranular”) films may play 
an important role in the development of emerging technologies. 
In particular, they show a resistive switching behavior \cite{lee2020}
that can be exploited in the fabrication of electrical devices able to
process and store data in the same physical unit
\cite{diventra2018,ielmini2018,traversa2015},
as requested by the neuromorphic computing paradigm
\cite{avizienis2012neuromorphic,7549034}.
Such behavior emerges in the near-to-percolation regime
\cite{borziak1976,sattar2013,minnai2017},
when deposited clusters do not form a complete layer of the film yet,
as well as when the film thickness is well above the electrical percolation threshold
\cite{mirigliano2019,mirigliano2020,mirigliano2020b,mirigliano2021a}.
In particular, for this latter situation a well-established explanation
of the underlying physical mechanisms is still missing.

Atomistic simulations may help to get insights on the microscopic mechanisms 
responsible for such phenomena. 
Correctly estimating the electrical conductivity becomes therefore crucial \cite{mirigliano2021a,tarantino2020}.
To this end, we developed an atomically informed mesoscopic model which provides accurate conductivity estimates 
for systems composed by interconnected gold nanoclusters.

The conductivity of nanogranular films is strongly affected by the high degree of porosity and defects in the metallic component. 
As the first step, we resort to atomistic simulations 
based on molecular dynamics to create realistic structures 
that capture the complexity of nanogranular films at the nanoscopic scale, 
a method that has been successfully used in the past to analyse morphology 
and mechanical properties of such systems \cite{benetti2017}. 
Provided with a realistic representation of the atomic-scale complexity of the system, 
we proceed by calculating its conductance by means of an Equivalent Resistor Network (ERN). Within such an approach, a system is typically approximated with a network of interconnected resistors whose impedance are determined by its local values; the overall resistance is therefore calculated using Kirchhoff’s circuit laws applied to the  network.

The typical scale of the inhomogeneities of a nanogranular film is, 
however, comparable to that of the electron mean free path, $l_e$, 
in the corresponding crystalline phase, \textit{i.e.} $l_e=37.5$ nm for Au \cite{gall2016electron}. 
Specializing on films with thickness well beyond the percolation threshold, 
we include in the ERN the ballistic component of electronic transport, 
which dominates at the length scale at which inhomogeneities occur, according to the following picture. 
Inhomogeneities in the metallic component of a nanogranular film 
are mainly due to the cluster landing impacts occurring during the deposition stage. 
They can be characterized as layers of highly disordered (amorphous) matter either between adjacent clusters 
or within the clusters themselves. We therefore model the metallic component as a collection 
of amorphous regions mixed with pristine crystalline ones and assume that electronic transport within each region and between regions of the same phase 
(whose length scale is typically just of a few nanometers) is mainly ballistic, 
while between regions of different phase the transport is diffusive. 
Such a picture is encoded in an ERN by requiring that resistors contribute 
to the overall resistance either ballistically or diffusively, 
whether they connect regions with same or different phases, respectively. 
While the diffusive-like behavior is readily obtained by letting 
the resistors abide by the classical Kirchhoff’s circuit laws, 
the ballistic-like behavior is enforced by assigning to each resistor 
a value of resistance that does not simply depend on the local structure of the system 
but on the entire region it belongs to
and reflects the size scaling typical of ballistic transport.

\begin{figure*}
\centering
\includegraphics[width=0.8\textwidth]{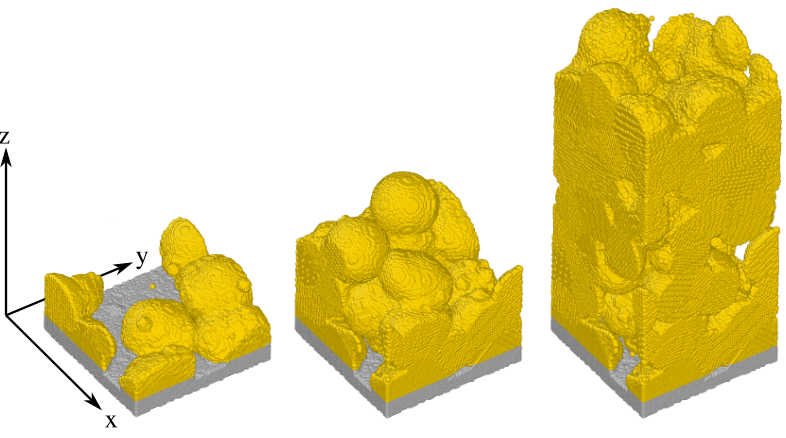}
\caption{View of the film growth at three different deposition stages: left panel shows just few Au clusters landed on the substrate (grey atoms). The corresponding thickness of the sample is $<6$ nm and no percolation path connecting two opposite sides of the sample has been created yet; in the middle panel the creation of the first percolation paths is already achieved, corresponding to a film thickness of $20$ nm; right panel shows the film in an advanced growth stage with a film thickness $>60$ nm.}
\label{fig:fig0}
\end{figure*}

Accurate characterizations of the ballistic component of electronic transport 
in metal nanostructures can be obtained using ab initio methods \cite{tavazza2011electron}.
In particular, we use density functional theory (DFT) combined 
with nonequilibrium Green’s function (NEGF) techniques 
(i) to study ballistic transport in structures mimicking the inhomogeneities found in the nanogranular film and (ii) to determine appropriate values of resistance for the ERN. 
Provided with such an input, the ERN can be finally used to get 
an estimate of the conductance of the entire simulated system.

To demonstrate the robustness of our procedure, 
we have simulated the growth of a nano-sized sample of a nanogranular film assembled 
by cluster deposition, close to the experimental conditions 
of Ref. \onlinecite{mirigliano2019}, 
and calculated its resistance at various stages of growth.

The paper is organized as follows: in Section \ref{section:met} we describe the methodology used to simulate the growth of the nanogranular Au film by means of classical Molecular Dynamics (MD). In Section \ref{section:trans} we discuss the procedure to accurately estimate the conductance of the specific gold micro-structures observed in the simulated film. In Section \ref{section:ERN} the ballistic ERN used to compute the total resistance of the film is presented. Finally, in Section \ref{section:Res} we present the results provided by our electrical model and compare them to a set of experimental results. 

\section{Simulated film growth and structural analysis}
\label{section:met}

The cluster assembled metallic film was obtained by simulating the multiple landing of 210 Au clusters deposited in 6 different steps \cite{benetti2017}, by classical MD. All the gold clusters were first thermalized  at 300 K for 150 ps. 
The size population of the clusters was constructed with 70\% of the clusters of diameters 8.8 nm and  30\% of diameters 1.3 nm, thus reproducing the  experimental size-distribution \cite{mirigliano2021a}.

As for the growth process, the first 35 Au clusters were deposited at random positions and normal impact direction on top of the substrate. The average kinetic energies per atom of the landing clusters were fixed to 0.25 eV/atom consistently with the results obtained from the experiments\cite{minnai2017facile}. Periodic boundary conditions were applied in the in-plane directions normal to the growth one. Finally, the clusters  were left free to evolve according to Newtonian dynamics. The snapshots of the film corresponding to three different deposition stages are displayed in Figure \ref{fig:fig0}: after the first deposition steps we observe a film characterized by isolated grains, most of the substrate surface being unoccupied. We can observe in the subsequent deposition steps the formation of cavities giving rise to the expected film porosity, \textit{i.e.} $\sim 30 \%$.

MD simulations have been performed using the LAMMPS code \cite{plimpton1993fast}, integrating the equations of motion by the velocity-Verlet algorithm.  The Nose-Hoover thermostat  with  relaxation time  equal  to  100  fs  was  used  to  control  the temperature.  
The Au-Au interactions were sampled using  a 12-6 Lennard-Jones potential with a cut off at 0.8 nm. The Lennard-Jones parameters have been optimized in order to reproduce several properties  such as  surface  tension  density  in  good  agreement  with  experiment\cite{heinz2008accurate}, \textit{i.e.} $\epsilon=5.29$ eV 
and $\sigma=2.62904$ \AA.
The Au substrate (grey atoms in Figure \ref{fig:fig0}) with dimensions of 24.5$\times$24.5$\times$5 nm$^3$ was constructed with the (111) surface exposed to the deposition of the clusters.  The four bottom layers were kept fixed in order to mimic a bulk material and a slab region (1.7 nm thick) adjacent to the fixed slab was thermalized at room temperature.

\begin{figure}
\centering
\includegraphics[width=0.5\textwidth]{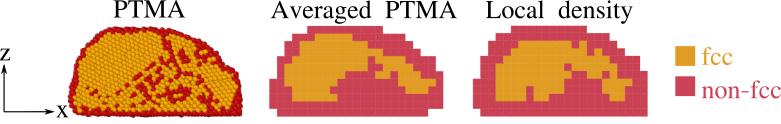}
\caption{$xz$ view of a single Au deposited cluster with radius $r=4.4$ nm after landing. The collision against the substrate strongly modifies its original spherical shape creating different micro-structural defects. Left panel: a Polyhedral Template Matching Analysis allows to localize with atomic resolution the distribution of the defects in the cluster. Orange atoms are found in fcc sites while red ones cannot be classified as such. Middle panel: the PTM analysis is averaged over the grid elements of the ERN allowing to distinguish fcc (orange) and non-fcc (red) regions. Right panel: distribution of fcc (orange) and non-fcc regions (red) based on the local density of the granule showing a good agreement with the averaged PTMA.}
\label{fig:fig1}
\end{figure}

Figure \ref{fig:fig1} displays a gold grain landed on the substrate: while before landing, by construction, the cluster is perfectly spherical, the collision with the substrate strongly affects its shape and structure. We observe in the deposited film two kinds of atomic arrangements: cubic (fcc) and non-cubic (non-fcc) gold. A Polyhedral Template Matching (PTM) Analysis performed with Ovito \cite{stukowski2009visualization}, a scientific analysis software for molecular simulation models, allows to distinguish between those Au atoms sitting in fcc sites and those which are not (related to planar and bulk defects). In the left panel of Figure \ref{fig:fig1} we show a section of the PTM analysis performed on the gold cluster after landing. For the sake of clarity we have excluded from the analysis those atoms belonging to the substrate. We observe that, due to the collision with the substrate, the fcc symmetry is broken, thus originating regions with different crystal structure. These defects are local and separate different fcc regions (orange colored atoms) within a single grain. In addition to that, the fcc symmetry is also broken at the surface of the cluster creating a non-fcc shell all around the grain. The shell is found before and after the collision, thus, it is not produced by the exceeding kinetic energy after landing. However, we do observe an increase of the shell thickness after cluster landing. Moreover, the non-fcc shell is responsible of having non-fcc layers separating adjacent deposited clusters in the film.

\begin{figure}
\centering
\includegraphics[width=0.4\textwidth]{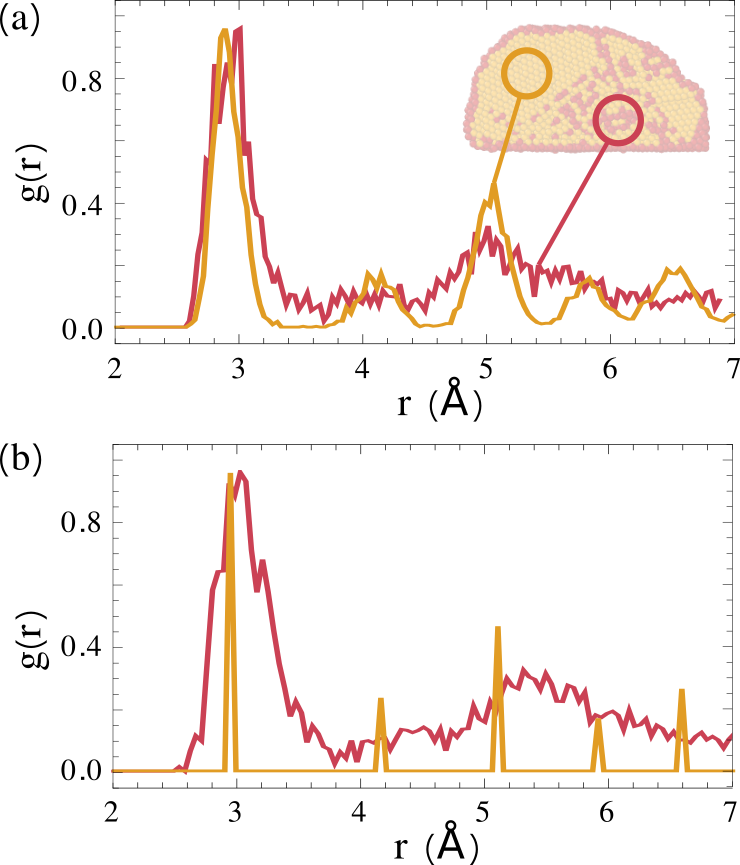}
\caption{Panel (a): Pair correlation function $g(r)$ computed on the highlighted regions of the cluster showing a first peak at $r=2.89$ \AA{} and $r=2.99$ \AA{} for fcc (orange) and non-fcc (red) phase, respectively. Panel (b): $g(r)$ for the DFT simulation cells showing a main peak placed at $r=2.95$ \AA{} and $r=3.04$ \AA{} for fcc and non-fcc, respectively. The cells are prepared to match the ratio obtained in the $g(r)$ distributions from the simulated sample showed in panel (a), \textit{i.e.} $b_{non-fcc}/b_{fcc}=1.03$.}
\label{fig:fig2}
\end{figure}

The PTM analysis provides very accurate atomically resolved structural phase maps of the clusters assembled to form the film. However, its heavy computational cost prevents from using PTM to distinguish the different phases during the evolution of the simulated film which counts with more than $10^6$ atoms. Therefore, we rather measure the local atomic density, $n_l$, of the film: we observe that the presence of defects induces a slight increase in the Au-Au bond length which shifts from $b_m=2.89$ \AA{} for fcc-Au regions to $b_m=2.99$ \AA{}  for non-fcc ones. This effect can be seen in Figure \ref{fig:fig2}(a) where the radial distribution function calculated on a sub-region of the granule containing fcc atoms (orange line) and non-fcc atoms (red line) is shown. A more evident effect is the broadening of the peaks for the non-fcc region. The upward shift of the first peak for non-fcc regions is directly translated to a decrease of the local density that we define as $n_l=N_r/V_r$, where $N_r$ is the number of gold atoms contained in that particular region of volume $V_r$. Two specific local density threshold values, $n_v$ and $n_c$, are used to set the density ranges corresponding to vacuum, non-fcc and fcc gold. We consider a region to be vacuum if its local density falls below $n_v$, while cubic gold is defined as $n_l>n_c$. Finally, non-cubic gold corresponds to $n_v<n_l<n_c$. In order to evaluate the agreement between the two methods we have averaged the former PTM analysis over the regions on which the local density approach is performed, \textit{i.e.} a regular grid with element size 5\AA$\times$5\AA$\times$5\AA, thus $V_r=0.125$ nm$^3$. By doing so we obtain the color map displayed in the middle panel of Figure \ref{fig:fig1}. In the right panel, the corresponding local density map is shown with $n_v=0.0005$ atoms/\AA$^3$ and $n_c=0.048$ atoms/\AA$^3$. Despite the much lower computational cost
a good agreement in the ratios between the fcc/non-fcc/vacuum occupied volume and the total volume is observed.
A side effect is a slight over-estimation of the non-fcc shell's thickness for the local density approach.

\section{electron transport in gold nano-junctions}
\label{section:trans}

The ballistic component is supposed to dominate
the electronic transport within individual nanoparticles at the considered length scales. 
Other transport mechanisms such as tunneling and hopping are not included since a strong-coupling regime \cite{beloborodov2007} is expected at the considered temperature and lengths, neither Coulomb interaction and quantum interference effects (see Supplemental Material).
We assume the two individuated gold phases, \textit{i.e.} fcc and non-fcc, to have different electronic transport characteristics. This is justified by the fact that non-fcc Au regions are characterized by a lack of symmetry that effectively reduces the number of opened conduction channels for ballistic transport \cite{kawamura1993quantum}. Moreover, we also assume that the estimation of the ballistic conductance of the two phases is sufficiently accurate to build up a reliable resistive model. 
The conductance of fcc and non-fcc Au is estimated by a blended NEGF-DFT approach.
In particular, the computation of the conductance of gold nano-sized systems \cite{dreher2005structure,kurui2009conductance,tavazza2011electron}
has been boosted by recent experiments on electronic properties 
of atomic-sized gold structures \cite{yanson2005atomic,oshima2003development,rodrigues2000signature,kiguchi2006conductance,suzuki2007distribution,kizuka2008atomic,yasuda1997conductance}.
The remarkable agreement between estimates and experimental values 
for different lengths, cross-sections and crystallographic orientations proves 
the accuracy of this approach in the study of electron transport 
in gold systems at the nano-scale.

\begin{figure}[t]
\centering
\includegraphics[width=0.45\textwidth]{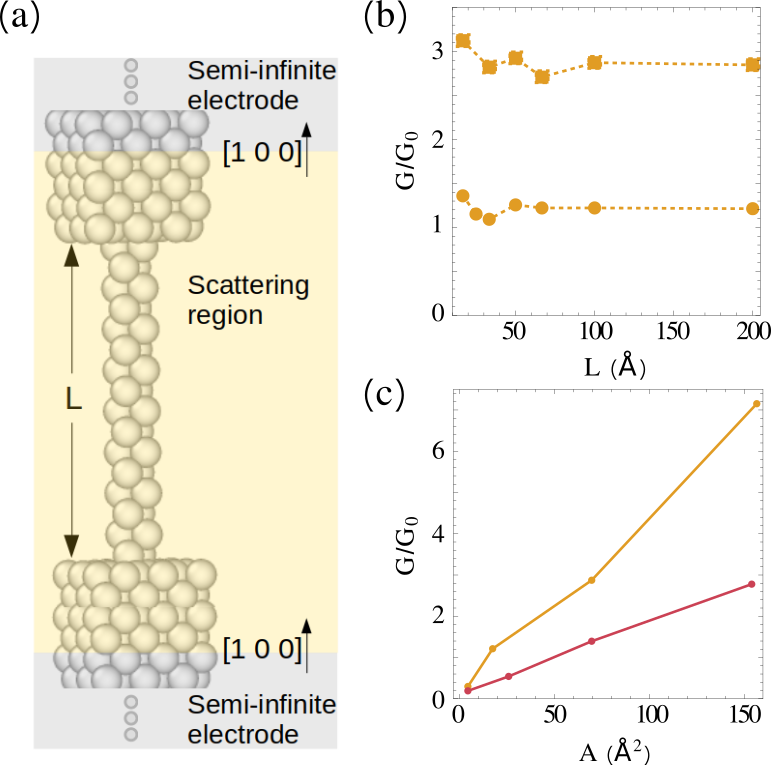}
\caption{Panel (a): schematic view of the simulated device. Two semi-infinite [100] Au electrodes (grey atoms) and the scattering region formed by a repetition of the electrodes plus the gold junction (yellow atoms). Panel (b): computed conductance $G$ for fcc-Au junctions with different lengths, $L$, for two different wire cross-sections, $A$: plateaus at 1.2 $G_0$ and 2.9 $G_0$ are obtained in agreement with Ref. \onlinecite{yanson2005atomic}. Panel (c): computed conductance, $G/G_0$, for different cross-section $A$ for fcc (orange line) and non-fcc (red line) Au junctions with $L=100$ \AA.}
\label{fig:fig3}
\end{figure}

Using such an approach, we are therefore able to perform a comprehensive study
of the conductivity of a two-terminal device containing gold junctions
mimicking the structures individuated in the structural analysis of the MD samples, \textit{i.e.} fcc and non-fcc. More specifically, in order to compute the conductance of gold junctions that might be representative of those found in the simulated film, we proceed as follows: we first set the crystallographic orientation of the device electrodes of the two-terminal device, once for all. Atomic-scale Au junctions are build up with different lengths, $L$, and cross-sections, $A$, in between the two electrodes. To mimic the fcc phase, we ask the atoms belonging to the central scattering region to keep the crystallographic orientation of the electrodes (as shown in \ref{fig:fig3}(a)), while non-fcc ones are requested to (i) present no specific crystallographic orientation under the PTMA and (ii) to present a broader radial distribution function than fcc as observed in \ref{fig:fig2}(a). This is equivalent to ask to each added atomic layer to change orientation with respect to the previous one. In such a scheme, the first allows electrons to see the symmetry of the lattice along the device, while the latter incorporates the non-homogeneity of the medium found in the deposited clusters. The $g(r)$ for fcc and non-fcc gold junctions are displayed in Figure \ref{fig:fig2}(b).
Another possible approach to evaluate the conductance of the defects individuated in the simulated film is to simply carve out from the film those regions we are interested in. The reason to avoid this approach is the fact that the non-homogeneities found in the film extend only for few atomic layers, while the required calculations that allow to specify a unique value for the conductance depending on $A$ and $L$ require the consideration of lengths and cross-sections beyond that limit (see length-scales in Figure \ref{fig:fig3}(b) and (c)).

We describe the gold electronic structure self-consistently using DFT within 
the Generalized Gradient Approximation (GGA) as implemented in the SIESTA package \cite{soler2002siesta}. Core electrons are modelled 
with Troullier–Martins nonlocal pseudopotentials, 
while the valence electrons are expanded with a double-$\zeta$ basis set. The mesh cutoff is $300$Ry and a 10$\times$10$\times$10 $k$-point mesh is used for the $4$ atoms unit-cell. We relax all the atomic coordinates till atomic forces are below 0.04 eV/\AA{}  and 0.10 eV/\AA{} for fcc and non-fcc Au, respectively.

For the conductance calculations, we have used TRANSIESTA \cite{brandbyge2002density}, which is based on the combination of DFT with the NEGF technique. Therefore, calculations on transport properties are based on the Landauer scheme of elastic scattering probability \cite{landauer1978}. Within such a scheme, given a certain bias, $V$, it is possible to compute the current, $I$, after self-consistently solving the NEGF and the electrostatic potential to get the electronic density matrix. The conductance of the device is then computed as $G=I/V$.
Semi-infinite $8$x$8$ $100$-Au electrodes, $4$ layers thick, sampled with a converged $k$-point grid of $3$x$3$x$20$, are considered. 

In the ballistic transport regime the conductance of a material is well known to be independent of the device length. The first step is to compute the conductance, $G$, for fcc gold junctions for different cross-sections and increasing lengths ranging from tens to hundreds of \AA. The computed values of $G$ against the wire length, $L$, are displayed in Figure \ref{fig:fig3}(b) in units of the quantum of conductance, $G_0=0.0000775$ $\Omega^{-1}$. We observe fluctuations in the computed $G$ values for short wires ($L<50$ \AA) while $G$ converges to a constant value for longer wires, as expected for ballistic transport.
Many experiments \cite{rodrigues2000signature,erts2000maxwell} and computational works \cite{dreher2005structure,kurui2009conductance} have reported an increase of $G$ for wires with increasing $A$, due to the increase in the number of opened conduction channels. A linear relation between $G$ and $A$ is expected with a slope depending on the crystallographic orientation \cite{yanson2005atomic}. We obtain values close to 1.2 $G_0$ and 2.9 $G_0$ for the two type of gold junctions considered in Figure \ref{fig:fig3}(b). In particular, as the cross-section of a wire oriented along the [1 0 0] direction is increased from $1.7$ \AA{} (corresponding to 4 unit cells) to $7.0$ \AA{} (corresponding to 16 unit cells) a increase of $\Delta G=$1.7 $G_0$ is observed corresponding to the transition from 1.2 $G_0$ to 2.9 $G_0$, close to the expected value provided by the simplified free electron model used in Ref. \onlinecite{yanson2005atomic}, \textit{i.e.} 1.8 $G_0$.
Figure \ref{fig:fig3}(c) shows the conductance for long wires ($L=100$ \AA) for increasing cross-sections for fcc (orange) and non-fcc Au wires (red).
The linear trend for fcc junctions is characterized by a slope of 0.04 $G_0/$\AA$^2$ while it is less pronounced for non-fcc, \textit{i.e.} 0.02 $G_0/$\AA$^2$. As expected, non-fcc Au junctions reveal less conductive than fcc ones, for all considered $A$.

\section{Equivalent Resistor Network with ballistic transport}
\label{section:ERN}

\begin{figure}
\centering
\includegraphics[width=0.47\textwidth]{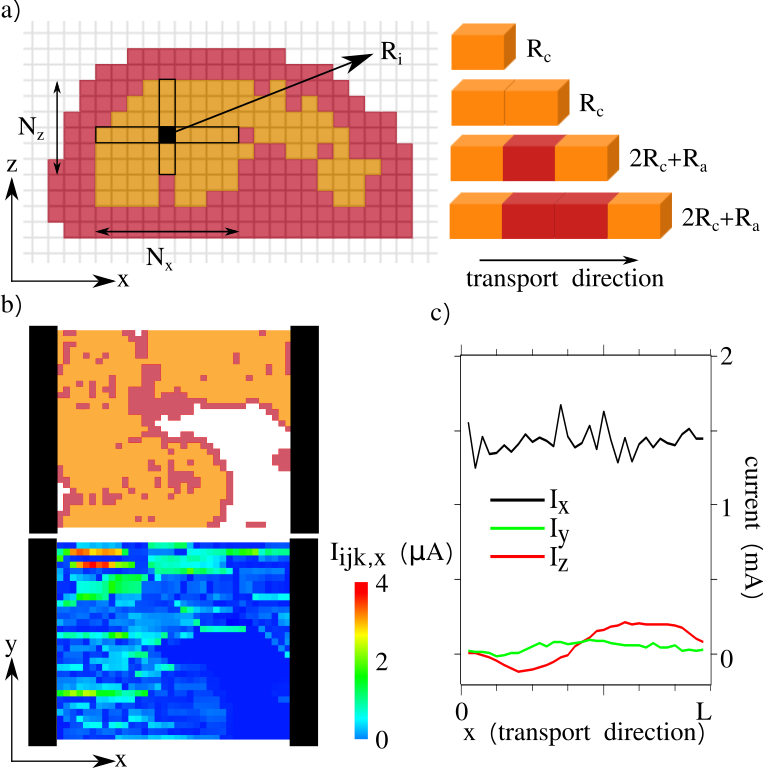}
\caption{Panel (a): (Left) $xz$ view of the regular 3D grid applied to a single gold cluster. For a given a mesh element, N$_i$ is the number of adjacent cells with equal gold phase (fcc or non-fcc) along the $i$-direction. (Right) Total resistance along a given transport direction for different combinations of fcc and non-fcc cells. R$_c$ and R$_a$ stand for the resistance value of a single fcc and non-fcc element, respectively. Panel (b): $xy$ view of fcc and non-fcc cells of the film at $z=15$ nm and the corresponding $I_{ijk,x}$ map ($k=30$) obtained from the converged solution of the ERN. Black strips represent the electrodes. Panel (c): the three components of the current vector, $(I_x,I_y,I_z)$ computed along the transport direction.}
\label{fig:fig5b}
\end{figure}

An ERN model is used to evaluate the electrical conductivity of the simulated film. A 3D regular grid is superimposed to the deposited film and each ERN cell is filled with one of the following: fcc Au, non-fcc Au or vacuum depending on the local value of $n_l$ as explained in Section \ref{section:met}. A $xy$ projection of the grid over a single landed gold cluster is shown in Figure \ref{fig:fig5b}(a). We assign to each ERN grid element a resistance value as follows: for vacuum elements this is set to $10^{15} \Omega$, while for Au elements, this is computed sticking to the geometrical dependencies found in Figure \ref{fig:fig3}(b) and (c). The rational behind the value chosen for vacuum is to assure those elements do not contribute to the final resistance value. This can be achieved by setting this value to infinite, which carries numerical issues. Instead, we choose to set this to a sufficiently high value with respect to the fcc and non-fcc ones, both falling in the k$\Omega$ range.

From the simulated sample we can roughly distinguish two types of interface: 
interfaces separating two grains with different crystallographic orientation (“fcc-fcc”) 
and those separating ordered grains from disordered regions (“fcc–non-fcc”). 
In our modelling, the latter type is always recognised as an interface that disrupts 
the electronic transport so the total resistance of a slab of fcc gold in contact 
with a non-fcc region is equal to the sum of their resistances, 
capturing the decoherence of electrons when reaching the interface and 
the interruption of the ballistic transport. In other words, 
the interface limits the regions where the electronic transport is considered ballistic. 
Instead, fcc-fcc interfaces, on the other hand, 
affect the transport only if the change of symmetry in going from one grain 
to the other is high enough so the density analysis detects the intermediation 
of a non-fcc region in between, otherwise the interface is effectively neglected. 
It must be remarked that clean fcc-fcc interfaces rarely occur 
as one can see in Figure \ref{fig:fig1}.

Thus, for a given mesh element containing Au atoms in a fcc (non-fcc) phase, we compute the number, (N${_x}$, N${_y}$, N${_z}$), of consecutive cells along each cartesian direction containing cubic (non-cubic) gold. We then give a unique resistance value for each transport direction as $R_i=R_b/N_i$ where $R_b=[G(A)]^{-1}$ and  $i=x, y, z$. Doing so we assure the total resistance of a given chunk of gold does not depend on its length, and only the cross-section $A$ determines its final value as expected for ballistic transport. Therefore, given a transport direction, if the total resistance value for a single fcc(non-fcc) mesh element is $R_c$($R_a$), the total resistance for N consecutive fcc(non-fcc) cells along that transport direction equals $R_c$($R_a$). Instead, the alternative stacking of fcc and non-fcc elements result in a total resistance that equals the sum of the consitutive parts, as represented in Figure \ref{fig:fig5b}(a).

The ERN grid counts with I$\times$J$\times$K elements the dimensions of which have been chosen so as (i) to minimize the computational cost of solving iteratively the ERN for the considered structures, and (ii) to have enough spatial resolution to well resolve the intra- and inter-granules structure. We set the element size to $5$ \AA{}, so the element volume and the minimum resolved area are $0.125$ nm$^3$ and $0.25$ nm$^2$, respectively. With this the ERN has 50$\times$50$\times$K elements with K increasing at each deposition step in order to include all deposited clusters.
Once all the mesh elements count with a resistance value, a finite bias, $V_{bias}$, across the sample is applied: the voltage is set to $V=V_{bias}$ for those grid elements belonging to one of the electrodes, while it is set to $V=0$ V otherwise. We have used $V_{bias}=0.06$ V to generate all the data included in Section \ref{section:Res}. We stress at this point that the $R_T$ estimation of the total film resistance produced by the present linear model does not depend on this parameter.

The obtained electrical network is analyzed by solving the Kirchhoff
equations. We solve them iteratively updating the node voltages
$V_{ijk}$ using the
formula
\begin{widetext}
\begin{equation}
V_{i j k}=\frac{\frac{V_{i-1,j,k}}{R_{(i-1) j
k,x}}+\frac{V_{i+1,j,k}}{R_{i j k,x}}+\frac{V_{i,j-1,k}}{R_{i (j-1)
k,y}}+\frac{V_{i,j+1,k}}{R_{i j k,y}}+\frac{V_{i,j,k-1}}{R_{i j
(k-1),z}}+\frac{V_{i,j,k+1}}{R_{i j
    k,z}}}{\frac{1}{R_{(i-1) j k,x}}+\frac{1}{R_{i j k,x}}+\frac{1}{R_{i
(j-1) k,y}}+\frac{1}{R_{i j k,y}}+\frac{1}{R_{i j k,z}}+\frac{1}{R_{i j
(k-1),z}}}
\end{equation}
\end{widetext}
where $R_{ijk,x}$ is the resistance of the $(i,j,k)$ grid element in the
$x$ direction, etc., and keeping fixed the electrode voltage. Iterations
are performed until the variation of the sample total resistance between
iteration steps is less than 0.01 $\Omega$.

From the node voltages and the resistances, one can calculate the
current flowing through the simulated sample. Each $(i,j,k)$ grid
element counts with a three component current vector $\{I_{ijk,x},I_{ijk,y},I_{ijk,z}\}$. In Figure \ref{fig:fig5b}(b) we show the converged current map at $z=15$ nm ($k=30$) for the simulated sample at a very advanced growth stage along with the corresponding fcc/non-fcc grid at that film height. The total current vector along the bias direction ($x$ direction in \ref{fig:fig5b}(b)) is computed as $\textbf{I}= (\sum_{jk} I_{ijk,x},\sum_{jk} I_{ijk,y},\sum_{jk} I_{ijk,z})$. The three components of the total current vector are displayed in \ref{fig:fig5b}(c). We observe that, after reaching the converged solution of the ERN, the $x$ component of the total current vector, $I_x$, fluctuates around a constant value (1.5 mA with $V_{bias}=0.06V$) all along the transport direction (black line in Right panel of Figure \ref{fig:fig5b}(c)). For $I_y$ and $I_z$ we observe the generation of internal currents that cancel each other (see $\sim 0$ values close to the extremes of both curves) so no current gets in or out the system through the directions normal to the transport one. 
Finally, we compute the total resistance of the film as $R_T=V_{bias}/\bar{I}$ where $\bar{I}$ is the mean value the total current flowing through the film along the bias direction.

To reach the converged solution for the ERN represents the most expensive part of the present model in terms of CPU time (600 seconds for a $50 \times 50 \times 75$ ERN run in a single core). 

\section{Results}
\label{section:Res}

\begin{figure}[t]
\centering
\includegraphics[width=0.45\textwidth]{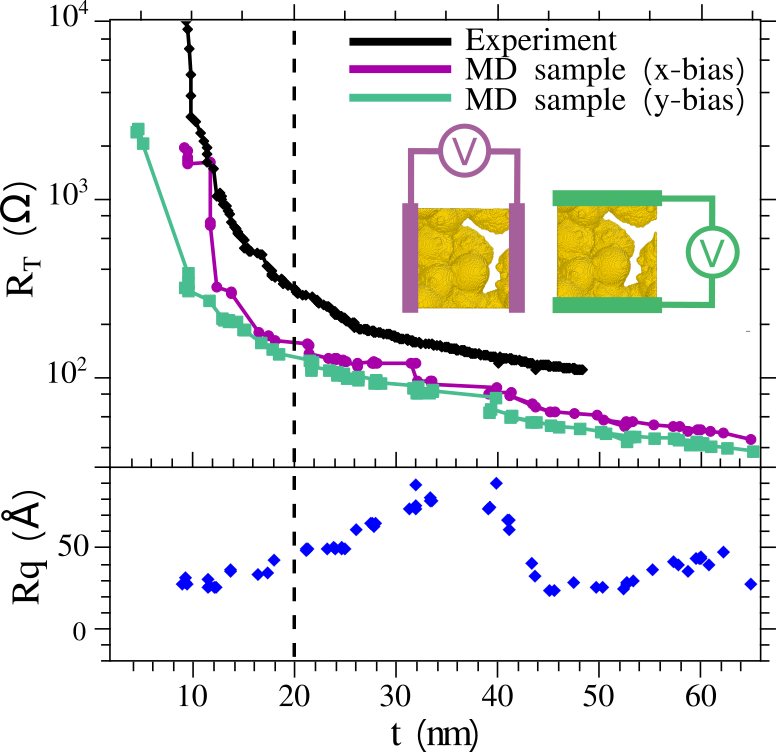}
\caption{Upper panel: percolation curves for the electrical resistance of cluster-assembled ﬁlms as function of the ﬁlm thickness on the x-axis (semi-log scale). Lower panel: evolution of the surface film roughness, $Rq$, for the simulated sample. The dashed line indicates the $t_m$ thickness at which $R_Tt^2$ reaches its minimum.}
\label{fig:fig5}
\end{figure}

\begin{figure}[t]
\centering
\includegraphics[width=0.45\textwidth]{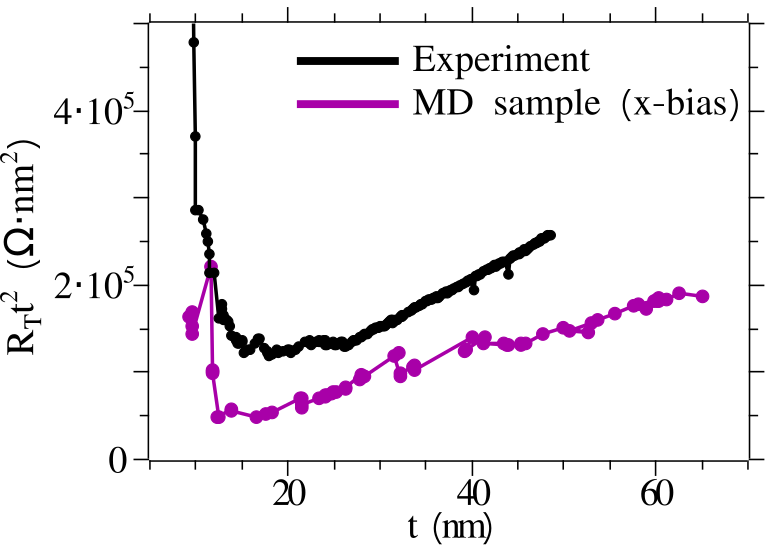}
\caption{$R_Tt^2$ as a function of film thickness $t$. The
minimum position on the simulated curve is determined with a parabolic fit to be at 20 nm. Black dots correspond to the experimental data and magenta dots to the simulated sample ($x$-bias)}.
\label{fig:fig6}
\end{figure}

The hierarchy of MD, NEGF-DFT and ERN models allows to compute the evolution of the film electrical resistance as gold clusters are deposited on the substrate. We have reproduced the percolation curves for the simulated sample and the corresponding data are displayed along with the experimental measurements from Ref. \onlinecite{mirigliano2019} in Figure \ref{fig:fig5}. We considered two different electrostatic bias conditions by setting the electrodes along the $x$ ($x$-bias) and $y$ directions ($y$-bias), as depicted in Figure \ref{fig:fig5}. The determination of the film thickness, $t$, is performed following the definition used to plot the results of the
experiments in Ref. \onlinecite{mirigliano2019}: we count the number of cells containing Au atoms in the ERN grid, $N_c$. Given the sample in-plane dimensions we compute $t_b$=0.125nm$^3N_c$/(24.5nm)$^2$ corresponding to the thickness of a bulk film with the same amount of matter. Finally, the porosity of the sample is introduced to obtain the final thickness value, $t=1.35t_b$.
We name $t_p$ the film thickness at which the first percolation path is created. For $t<t_p$, the film is characterized by isolated clusters and a infinite resistance ($>10^{13}$ $\Omega$ for numerical reasons). 
Since the current implementation does not take into account electron tunneling and hopping effects, our data starts being meaningful after the creation of the first percolation path, which creates a real contact between the the electrodes. 
We observe the first percolation path occurring at $t_{p,x}=5$ nm and $t_{p,y}=10$ nm, for the two considered bias conditions. After that, two growth stages are identified from the resistance-thickness curve \cite{stauffer1994}. At first, few inter-grain electrical contacts exist producing a film characterized by poorly connected aggregates and $R_T$ values in the 1-10 k$\Omega$ range. In this stage, known as geometrical percolation stage, $R_T$ abruptly decreases down to hundreds of $\Omega$ due to the increase in the paths becoming available for electron transport as clusters land and interconnect.
Next, a transition from insulating to ohmic behavior is observed, i.e $R_T$ smoothly decreases as $t$ is increased. This transition is defined as the thickness at which the quantity $R_Tt^2$ reaches its minimum value $t_m$ \cite{maaroof1994onset,burgmann2005electrical}. The good agreement achieved for the determination of this parameter is shown in Figure \ref{fig:fig6}. For all data sets the transition to ohmic behavior is achieved around $t_m=20$ nm. However, this parameter can suffer huge fluctuations from one sample to another \cite{mirigliano2019}. 
The predicted $R_T$ values show larger fluctuations in comparison to the experimental curve, these being more important at the first percolation steps. The fact that the simulated film has nano-scale dimensions, \textit{i.e.} 25$\times$25nm${^2}$, make granular effects still visible at all deposition stages and the landing of a single cluster produces huge variations in the computed values. Instead, the experimental curve corresponding to a macro-scale sample, presents a smoother behavior.

For $t>t_m$, the total film resistance further decreases (see Figure \ref{fig:fig5}). Quantitatively, our electrical model slightly underestimates the total resistance of the film for every deposition step. For instance, the ratio between the computed and the measured value at $t=40$ nm is $0.6$ and $0.7$ for $x$-bias and $y$-bias, respectively. Both experimental and simulated data follow a power law decay ($R_T\propto1/t^{\alpha}$) as clusters are deposited and this represents a qualitative agreement with the experimental data set. The experimental curve provides $\alpha_{exp}=0.9$, implying that the resulting film resistivity, calculated as $\rho=R_Tt$, should increase instead of remaining constant as expected for bulk materials and also observed for atomically assembled Au films. The power-law exponents for the two simulated curves equal to $\alpha_x=0.94$ and $\alpha_y=0.95$ for that thickness range, fulfilling the condition $\alpha<1$ to observe an increase of $\rho$. Indeed a quite remarkable agreement with experimental results.

\begin{figure}[t]
\centering
\includegraphics[width=0.4\textwidth]{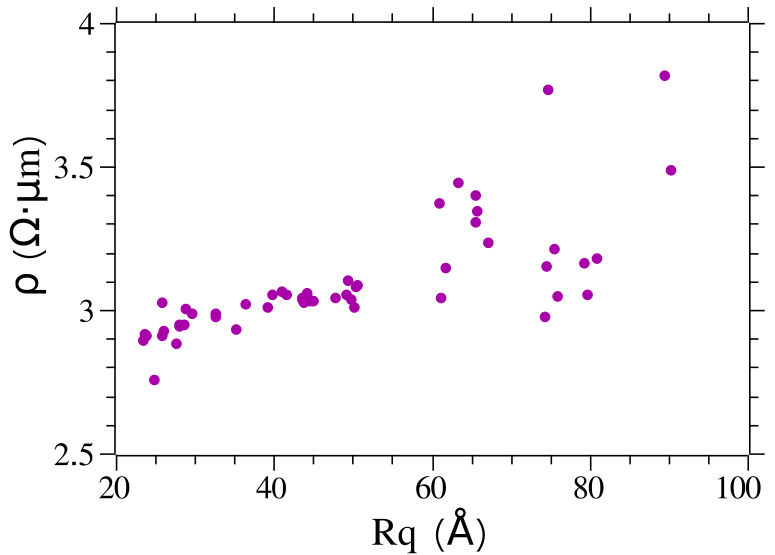}
\caption{$\rho$ vs. $Rq$ curve for the simulated sample for $t>20$ nm.}
\label{fig:fig7}
\end{figure}

This effect is often attributed to an increase of the sample roughness as suggested in Ref. \onlinecite{barborini2010influence,Namba_1970,munoz2017size}. In order to support the plausibility of this hypothesis we have computed the roughness of the film surface, $Rq$, which we estimate by means of the standard deviation of the film height measured on every $(i,j)$ mesh element as $Rq=\sqrt{\frac{\sum^{N}(z_{ij}-\bar{z})^2}{N-1}}$ where $z_{ij}$ is the height profile of the sample and $\bar{z}$ stands for the mean sample height. Differently to experimental realizations, we do have access to $Rq$ at every deposition step so a sound characterization of the film roughness dynamics is achieved. In Figure \ref{fig:fig5}(b) the evolution of $Rq$ is shown, displaying clearly two alternate regimes: one in which clusters are accumulated in few spots of the $(x,y)$ plane producing an increase of $Rq$ corresponding to $t<40$ nm and $t>45$ nm, and another in which the deposition of few clusters are enough to drastically reduce $Rq$ generating void regions within the film which results in the expected film porosity. The former, characterized by a non-uniform deposition of clusters, produces a reduction of the slope of the $R_T$ curve, due to the fact that the clusters do not contribute to create new conduction paths. The latter, in which new clusters occupy regions connecting separated gold regions, effectively reduces the film resistance.
The periods of these two regimes are expected to depend on the ratio between the sample size and the clusters dimensions, being shorter for smaller ones. The correlation between the modeled film resistivity and its roughness for $t>40$ nm can be observed in Figure \ref{fig:fig7}. Thus, the present electrical model well captures the role of surface roughness in the evolution of the film growth.\\

\section{Conclusions}
\label{section:concl}
We have presented an electrical model able to predict the resistance of cluster-assembled gold films based upon a well resolved description of their nanostructure and related charge transport at that length scales. By reproducing the deposition of tens of Au clusters on a substrate by means of MD simulations we are able to reproduce the growth of a Au nanogranular film. Next, by using our model at different growth stages we compute the resistance percolation curve and compare the model predictions against experimental values obtaining a good agreement. In particular, the power-law dependence of the total resistance with the film thickness is well reproduced as well as the non-monotonic behavior of the film resistivity and its dependence with the film surface roughness. The modeled film is a $\sim$30$\%$ more conductive respect to the experimental one, likely due to the simplified synopsis of transport mechanisms it is based on. Further improvement of the model could be achieved by including tunneling and hopping transport mechanisms so to extend the applicability of the model to the $t<t_p$ region of the percolation curve.

\begin{acknowledgments}
This work was fully funded by Fondazione CON IL SUD (Grant No: 2018-PDR-01004).
\end{acknowledgments}


\begin{thebibliography}{41}%
\makeatletter
\providecommand \@ifxundefined [1]{%
 \@ifx{#1\undefined}
}%
\providecommand \@ifnum [1]{%
 \ifnum #1\expandafter \@firstoftwo
 \else \expandafter \@secondoftwo
 \fi
}%
\providecommand \@ifx [1]{%
 \ifx #1\expandafter \@firstoftwo
 \else \expandafter \@secondoftwo
 \fi
}%
\providecommand \natexlab [1]{#1}%
\providecommand \enquote  [1]{``#1''}%
\providecommand \bibnamefont  [1]{#1}%
\providecommand \bibfnamefont [1]{#1}%
\providecommand \citenamefont [1]{#1}%
\providecommand \href@noop [0]{\@secondoftwo}%
\providecommand \href [0]{\begingroup \@sanitize@url \@href}%
\providecommand \@href[1]{\@@startlink{#1}\@@href}%
\providecommand \@@href[1]{\endgroup#1\@@endlink}%
\providecommand \@sanitize@url [0]{\catcode `\\12\catcode `\$12\catcode
  `\&12\catcode `\#12\catcode `\^12\catcode `\_12\catcode `\%12\relax}%
\providecommand \@@startlink[1]{}%
\providecommand \@@endlink[0]{}%
\providecommand \url  [0]{\begingroup\@sanitize@url \@url }%
\providecommand \@url [1]{\endgroup\@href {#1}{\urlprefix }}%
\providecommand \urlprefix  [0]{URL }%
\providecommand \Eprint [0]{\href }%
\providecommand \doibase [0]{https://doi.org/}%
\providecommand \selectlanguage [0]{\@gobble}%
\providecommand \bibinfo  [0]{\@secondoftwo}%
\providecommand \bibfield  [0]{\@secondoftwo}%
\providecommand \translation [1]{[#1]}%
\providecommand \BibitemOpen [0]{}%
\providecommand \bibitemStop [0]{}%
\providecommand \bibitemNoStop [0]{.\EOS\space}%
\providecommand \EOS [0]{\spacefactor3000\relax}%
\providecommand \BibitemShut  [1]{\csname bibitem#1\endcsname}%
\let\auto@bib@innerbib\@empty
\bibitem [{\citenamefont {Lee}\ \emph {et~al.}(2020)\citenamefont {Lee},
  \citenamefont {Zhu},\ and\ \citenamefont {Lu}}]{lee2020}%
  \BibitemOpen
  \bibfield  {author} {\bibinfo {author} {\bibfnamefont {S.~H.}\ \bibnamefont
  {Lee}}, \bibinfo {author} {\bibfnamefont {X.}~\bibnamefont {Zhu}},\ and\
  \bibinfo {author} {\bibfnamefont {W.~D.}\ \bibnamefont {Lu}},\ }\bibfield
  {title} {\bibinfo {title} {Nanoscale resistive switching devices for memory
  and computing applications},\ }\bibfield  {journal} {\bibinfo  {journal}
  {Nano Research}\ }\href {https://doi.org/10.1007/s12274-020-2616-0}
  {10.1007/s12274-020-2616-0} (\bibinfo {year} {2020})\BibitemShut {NoStop}%
\bibitem [{\citenamefont {Di~Ventra}\ and\ \citenamefont
  {Traversa}(2018)}]{diventra2018}%
  \BibitemOpen
  \bibfield  {author} {\bibinfo {author} {\bibfnamefont {M.}~\bibnamefont
  {Di~Ventra}}\ and\ \bibinfo {author} {\bibfnamefont {F.~L.}\ \bibnamefont
  {Traversa}},\ }\bibfield  {title} {\bibinfo {title} {Perspective:
  Memcomputing: Leveraging memory and physics to compute efficiently},\ }\href
  {https://doi.org/10.1063/1.5026506} {\bibfield  {journal} {\bibinfo
  {journal} {Journal of Applied Physics}\ }\textbf {\bibinfo {volume} {123}},\
  \bibinfo {pages} {180901} (\bibinfo {year} {2018})},\ \Eprint
  {https://arxiv.org/abs/https://doi.org/10.1063/1.5026506}
  {https://doi.org/10.1063/1.5026506} \BibitemShut {NoStop}%
\bibitem [{\citenamefont {Ielmini}\ and\ \citenamefont
  {Wong}(2018)}]{ielmini2018}%
  \BibitemOpen
  \bibfield  {author} {\bibinfo {author} {\bibfnamefont {D.}~\bibnamefont
  {Ielmini}}\ and\ \bibinfo {author} {\bibfnamefont {H.-S.~P.}\ \bibnamefont
  {Wong}},\ }\bibfield  {title} {\bibinfo {title} {In-memory computing with
  resistive switching devices},\ }\href
  {https://doi.org/10.1038/s41928-018-0092-2} {\bibfield  {journal} {\bibinfo
  {journal} {Nature Electronics}\ }\textbf {\bibinfo {volume} {1}},\ \bibinfo
  {pages} {333} (\bibinfo {year} {2018})}\BibitemShut {NoStop}%
\bibitem [{\citenamefont {Traversa}\ and\ \citenamefont
  {Di~Ventra}(2015)}]{traversa2015}%
  \BibitemOpen
  \bibfield  {author} {\bibinfo {author} {\bibfnamefont {F.~L.}\ \bibnamefont
  {Traversa}}\ and\ \bibinfo {author} {\bibfnamefont {M.}~\bibnamefont
  {Di~Ventra}},\ }\bibfield  {title} {\bibinfo {title} {Universal memcomputing
  machines},\ }\href {https://doi.org/10.1109/TNNLS.2015.2391182} {\bibfield
  {journal} {\bibinfo  {journal} {IEEE Transactions on Neural Networks and
  Learning Systems}\ }\textbf {\bibinfo {volume} {26}},\ \bibinfo {pages}
  {2702} (\bibinfo {year} {2015})}\BibitemShut {NoStop}%
\bibitem [{\citenamefont {Avizienis}\ \emph {et~al.}(2012)\citenamefont
  {Avizienis}, \citenamefont {Sillin}, \citenamefont {Martin-Olmos},
  \citenamefont {Shieh}, \citenamefont {Aono}, \citenamefont {Stieg},\ and\
  \citenamefont {Gimzewski}}]{avizienis2012neuromorphic}%
  \BibitemOpen
  \bibfield  {author} {\bibinfo {author} {\bibfnamefont {A.~V.}\ \bibnamefont
  {Avizienis}}, \bibinfo {author} {\bibfnamefont {H.~O.}\ \bibnamefont
  {Sillin}}, \bibinfo {author} {\bibfnamefont {C.}~\bibnamefont
  {Martin-Olmos}}, \bibinfo {author} {\bibfnamefont {H.~H.}\ \bibnamefont
  {Shieh}}, \bibinfo {author} {\bibfnamefont {M.}~\bibnamefont {Aono}},
  \bibinfo {author} {\bibfnamefont {A.~Z.}\ \bibnamefont {Stieg}},\ and\
  \bibinfo {author} {\bibfnamefont {J.~K.}\ \bibnamefont {Gimzewski}},\
  }\bibfield  {title} {\bibinfo {title} {Neuromorphic atomic switch networks},\
  }\href@noop {} {\bibfield  {journal} {\bibinfo  {journal} {PloS one}\
  }\textbf {\bibinfo {volume} {7}},\ \bibinfo {pages} {e42772} (\bibinfo {year}
  {2012})}\BibitemShut {NoStop}%
\bibitem [{\citenamefont {Nawrocki}\ \emph {et~al.}(2016)\citenamefont
  {Nawrocki}, \citenamefont {Voyles},\ and\ \citenamefont {Shaheen}}]{7549034}%
  \BibitemOpen
  \bibfield  {author} {\bibinfo {author} {\bibfnamefont {R.~A.}\ \bibnamefont
  {Nawrocki}}, \bibinfo {author} {\bibfnamefont {R.~M.}\ \bibnamefont
  {Voyles}},\ and\ \bibinfo {author} {\bibfnamefont {S.~E.}\ \bibnamefont
  {Shaheen}},\ }\bibfield  {title} {\bibinfo {title} {A mini review of
  neuromorphic architectures and implementations},\ }\href
  {https://doi.org/10.1109/TED.2016.2598413} {\bibfield  {journal} {\bibinfo
  {journal} {IEEE Transactions on Electron Devices}\ }\textbf {\bibinfo
  {volume} {63}},\ \bibinfo {pages} {3819} (\bibinfo {year}
  {2016})}\BibitemShut {NoStop}%
\bibitem [{\citenamefont {Borziak}\ \emph {et~al.}(1976)\citenamefont
  {Borziak}, \citenamefont {Dyukov}, \citenamefont {Kostenko}, \citenamefont
  {Kulyupin},\ and\ \citenamefont {Nepijko}}]{borziak1976}%
  \BibitemOpen
  \bibfield  {author} {\bibinfo {author} {\bibfnamefont {P.}~\bibnamefont
  {Borziak}}, \bibinfo {author} {\bibfnamefont {V.}~\bibnamefont {Dyukov}},
  \bibinfo {author} {\bibfnamefont {A.}~\bibnamefont {Kostenko}}, \bibinfo
  {author} {\bibfnamefont {Y.}~\bibnamefont {Kulyupin}},\ and\ \bibinfo
  {author} {\bibfnamefont {S.}~\bibnamefont {Nepijko}},\ }\bibfield  {title}
  {\bibinfo {title} {Electrical conductivity in structurally inhomogeneous
  discontinuous metal films},\ }\href
  {https://doi.org/https://doi.org/10.1016/0040-6090(76)90389-8} {\bibfield
  {journal} {\bibinfo  {journal} {Thin Solid Films}\ }\textbf {\bibinfo
  {volume} {36}},\ \bibinfo {pages} {21 } (\bibinfo {year} {1976})}\BibitemShut
  {NoStop}%
\bibitem [{\citenamefont {Sattar}\ \emph {et~al.}(2013)\citenamefont {Sattar},
  \citenamefont {Fostner},\ and\ \citenamefont {Brown}}]{sattar2013}%
  \BibitemOpen
  \bibfield  {author} {\bibinfo {author} {\bibfnamefont {A.}~\bibnamefont
  {Sattar}}, \bibinfo {author} {\bibfnamefont {S.}~\bibnamefont {Fostner}},\
  and\ \bibinfo {author} {\bibfnamefont {S.~A.}\ \bibnamefont {Brown}},\
  }\bibfield  {title} {\bibinfo {title} {Quantized conductance and switching in
  percolating nanoparticle films},\ }\href
  {https://doi.org/10.1103/PhysRevLett.111.136808} {\bibfield  {journal}
  {\bibinfo  {journal} {Phys. Rev. Lett.}\ }\textbf {\bibinfo {volume} {111}},\
  \bibinfo {pages} {136808} (\bibinfo {year} {2013})}\BibitemShut {NoStop}%
\bibitem [{\citenamefont {Minnai}\ \emph
  {et~al.}(2017{\natexlab{a}})\citenamefont {Minnai}, \citenamefont
  {Bellacicca}, \citenamefont {Brown},\ and\ \citenamefont
  {Milani}}]{minnai2017}%
  \BibitemOpen
  \bibfield  {author} {\bibinfo {author} {\bibfnamefont {C.}~\bibnamefont
  {Minnai}}, \bibinfo {author} {\bibfnamefont {A.}~\bibnamefont {Bellacicca}},
  \bibinfo {author} {\bibfnamefont {S.~A.}\ \bibnamefont {Brown}},\ and\
  \bibinfo {author} {\bibfnamefont {P.}~\bibnamefont {Milani}},\ }\bibfield
  {title} {\bibinfo {title} {Facile fabrication of complex networks of
  memristive devices},\ }\href {https://doi.org/10.1038/s41598-017-08244-y}
  {\bibfield  {journal} {\bibinfo  {journal} {Scientific Reports}\ }\textbf
  {\bibinfo {volume} {7}},\ \bibinfo {pages} {7955} (\bibinfo {year}
  {2017}{\natexlab{a}})}\BibitemShut {NoStop}%
\bibitem [{\citenamefont {Mirigliano}\ \emph {et~al.}(2019)\citenamefont
  {Mirigliano}, \citenamefont {Borghi}, \citenamefont {Podestà}, \citenamefont
  {Antidormi}, \citenamefont {Colombo},\ and\ \citenamefont
  {Milani}}]{mirigliano2019}%
  \BibitemOpen
  \bibfield  {author} {\bibinfo {author} {\bibfnamefont {M.}~\bibnamefont
  {Mirigliano}}, \bibinfo {author} {\bibfnamefont {F.}~\bibnamefont {Borghi}},
  \bibinfo {author} {\bibfnamefont {A.}~\bibnamefont {Podestà}}, \bibinfo
  {author} {\bibfnamefont {A.}~\bibnamefont {Antidormi}}, \bibinfo {author}
  {\bibfnamefont {L.}~\bibnamefont {Colombo}},\ and\ \bibinfo {author}
  {\bibfnamefont {P.}~\bibnamefont {Milani}},\ }\bibfield  {title} {\bibinfo
  {title} {Non-ohmic behavior and resistive switching of au cluster-assembled
  films beyond the percolation threshold},\ }\href
  {https://doi.org/10.1039/C9NA00256A} {\bibfield  {journal} {\bibinfo
  {journal} {Nanoscale Adv.}\ }\textbf {\bibinfo {volume} {1}},\ \bibinfo
  {pages} {3119} (\bibinfo {year} {2019})}\BibitemShut {NoStop}%
\bibitem [{\citenamefont {Mirigliano}\ \emph
  {et~al.}(2020{\natexlab{a}})\citenamefont {Mirigliano}, \citenamefont
  {Decastri}, \citenamefont {Pullia}, \citenamefont {Dellasega}, \citenamefont
  {Casu}, \citenamefont {Falqui},\ and\ \citenamefont
  {Milani}}]{mirigliano2020}%
  \BibitemOpen
  \bibfield  {author} {\bibinfo {author} {\bibfnamefont {M.}~\bibnamefont
  {Mirigliano}}, \bibinfo {author} {\bibfnamefont {D.}~\bibnamefont
  {Decastri}}, \bibinfo {author} {\bibfnamefont {A.}~\bibnamefont {Pullia}},
  \bibinfo {author} {\bibfnamefont {D.}~\bibnamefont {Dellasega}}, \bibinfo
  {author} {\bibfnamefont {A.}~\bibnamefont {Casu}}, \bibinfo {author}
  {\bibfnamefont {A.}~\bibnamefont {Falqui}},\ and\ \bibinfo {author}
  {\bibfnamefont {P.}~\bibnamefont {Milani}},\ }\bibfield  {title} {\bibinfo
  {title} {Complex electrical spiking activity in resistive switching
  nanostructured au two-terminal devices},\ }\href
  {https://doi.org/10.1088/1361-6528/ab76ec} {\bibfield  {journal} {\bibinfo
  {journal} {Nanotechnology}\ }\textbf {\bibinfo {volume} {31}},\ \bibinfo
  {pages} {234001} (\bibinfo {year} {2020}{\natexlab{a}})}\BibitemShut
  {NoStop}%
\bibitem [{\citenamefont {Mirigliano}\ \emph
  {et~al.}(2020{\natexlab{b}})\citenamefont {Mirigliano}, \citenamefont
  {Radice}, \citenamefont {Falqui}, \citenamefont {Casu}, \citenamefont
  {Cavaliere},\ and\ \citenamefont {Milani}}]{mirigliano2020b}%
  \BibitemOpen
  \bibfield  {author} {\bibinfo {author} {\bibfnamefont {M.}~\bibnamefont
  {Mirigliano}}, \bibinfo {author} {\bibfnamefont {S.}~\bibnamefont {Radice}},
  \bibinfo {author} {\bibfnamefont {A.}~\bibnamefont {Falqui}}, \bibinfo
  {author} {\bibfnamefont {A.}~\bibnamefont {Casu}}, \bibinfo {author}
  {\bibfnamefont {F.}~\bibnamefont {Cavaliere}},\ and\ \bibinfo {author}
  {\bibfnamefont {P.}~\bibnamefont {Milani}},\ }\bibfield  {title} {\bibinfo
  {title} {Anomalous electrical conduction and negative temperature coefficient
  of resistance in nanostructured gold resistive switching films},\ }\href
  {https://doi.org/10.1038/s41598-020-76632-y} {\bibfield  {journal} {\bibinfo
  {journal} {Scientific Reports}\ }\textbf {\bibinfo {volume} {10}},\ \bibinfo
  {pages} {19613} (\bibinfo {year} {2020}{\natexlab{b}})}\BibitemShut {NoStop}%
\bibitem [{\citenamefont {Mirigliano}\ and\ \citenamefont
  {Milani}(2021)}]{mirigliano2021a}%
  \BibitemOpen
  \bibfield  {author} {\bibinfo {author} {\bibfnamefont {M.}~\bibnamefont
  {Mirigliano}}\ and\ \bibinfo {author} {\bibfnamefont {P.}~\bibnamefont
  {Milani}},\ }\bibfield  {title} {\bibinfo {title} {Electrical conduction in
  nanogranular cluster-assembled metallic films},\ }\href
  {https://doi.org/10.1080/23746149.2021.1908847} {\bibfield  {journal}
  {\bibinfo  {journal} {Advances in Physics: X}\ }\textbf {\bibinfo {volume}
  {6}},\ \bibinfo {pages} {1908847} (\bibinfo {year} {2021})},\ \Eprint
  {https://arxiv.org/abs/https://doi.org/10.1080/23746149.2021.1908847}
  {https://doi.org/10.1080/23746149.2021.1908847} \BibitemShut {NoStop}%
\bibitem [{\citenamefont {Tarantino}\ and\ \citenamefont
  {Colombo}(2020)}]{tarantino2020}%
  \BibitemOpen
  \bibfield  {author} {\bibinfo {author} {\bibfnamefont {W.}~\bibnamefont
  {Tarantino}}\ and\ \bibinfo {author} {\bibfnamefont {L.}~\bibnamefont
  {Colombo}},\ }\bibfield  {title} {\bibinfo {title} {Modeling resistive
  switching in nanogranular metal films},\ }\href
  {https://doi.org/10.1103/PhysRevResearch.2.043389} {\bibfield  {journal}
  {\bibinfo  {journal} {Phys. Rev. Research}\ }\textbf {\bibinfo {volume}
  {2}},\ \bibinfo {pages} {043389} (\bibinfo {year} {2020})}\BibitemShut
  {NoStop}%
\bibitem [{\citenamefont {Benetti}\ \emph {et~al.}(2017)\citenamefont
  {Benetti}, \citenamefont {Caddeo}, \citenamefont {Melis}, \citenamefont
  {Ferrini}, \citenamefont {Giannetti}, \citenamefont {Winckelmans},
  \citenamefont {Bals}, \citenamefont {Van~Bael}, \citenamefont {Cavaliere},
  \citenamefont {Gavioli},\ and\ \citenamefont {Banfi}}]{benetti2017}%
  \BibitemOpen
  \bibfield  {author} {\bibinfo {author} {\bibfnamefont {G.}~\bibnamefont
  {Benetti}}, \bibinfo {author} {\bibfnamefont {C.}~\bibnamefont {Caddeo}},
  \bibinfo {author} {\bibfnamefont {C.}~\bibnamefont {Melis}}, \bibinfo
  {author} {\bibfnamefont {G.}~\bibnamefont {Ferrini}}, \bibinfo {author}
  {\bibfnamefont {C.}~\bibnamefont {Giannetti}}, \bibinfo {author}
  {\bibfnamefont {N.}~\bibnamefont {Winckelmans}}, \bibinfo {author}
  {\bibfnamefont {S.}~\bibnamefont {Bals}}, \bibinfo {author} {\bibfnamefont
  {M.~J.}\ \bibnamefont {Van~Bael}}, \bibinfo {author} {\bibfnamefont
  {E.}~\bibnamefont {Cavaliere}}, \bibinfo {author} {\bibfnamefont
  {L.}~\bibnamefont {Gavioli}},\ and\ \bibinfo {author} {\bibfnamefont
  {F.}~\bibnamefont {Banfi}},\ }\bibfield  {title} {\bibinfo {title} {Bottom-up
  mechanical nanometrology of granular ag nanoparticles thin films},\ }\href
  {https://doi.org/10.1021/acs.jpcc.7b05795} {\bibfield  {journal} {\bibinfo
  {journal} {The Journal of Physical Chemistry C}\ }\textbf {\bibinfo {volume}
  {121}},\ \bibinfo {pages} {22434} (\bibinfo {year} {2017})},\ \Eprint
  {https://arxiv.org/abs/https://doi.org/10.1021/acs.jpcc.7b05795}
  {https://doi.org/10.1021/acs.jpcc.7b05795} \BibitemShut {NoStop}%
\bibitem [{\citenamefont {Gall}(2016)}]{gall2016electron}%
  \BibitemOpen
  \bibfield  {author} {\bibinfo {author} {\bibfnamefont {D.}~\bibnamefont
  {Gall}},\ }\bibfield  {title} {\bibinfo {title} {Electron mean free path in
  elemental metals},\ }\href@noop {} {\bibfield  {journal} {\bibinfo  {journal}
  {Journal of Applied Physics}\ }\textbf {\bibinfo {volume} {119}},\ \bibinfo
  {pages} {085101} (\bibinfo {year} {2016})}\BibitemShut {NoStop}%
\bibitem [{\citenamefont {Tavazza}\ \emph {et~al.}(2011)\citenamefont
  {Tavazza}, \citenamefont {Smith}, \citenamefont {Levine}, \citenamefont
  {Pratt},\ and\ \citenamefont {Chaka}}]{tavazza2011electron}%
  \BibitemOpen
  \bibfield  {author} {\bibinfo {author} {\bibfnamefont {F.}~\bibnamefont
  {Tavazza}}, \bibinfo {author} {\bibfnamefont {D.~T.}\ \bibnamefont {Smith}},
  \bibinfo {author} {\bibfnamefont {L.~E.}\ \bibnamefont {Levine}}, \bibinfo
  {author} {\bibfnamefont {J.~R.}\ \bibnamefont {Pratt}},\ and\ \bibinfo
  {author} {\bibfnamefont {A.~M.}\ \bibnamefont {Chaka}},\ }\bibfield  {title}
  {\bibinfo {title} {Electron transport in gold nanowires: Stable 1-, 2-and
  3-dimensional atomic structures and noninteger conduction states},\
  }\href@noop {} {\bibfield  {journal} {\bibinfo  {journal} {Physical review
  letters}\ }\textbf {\bibinfo {volume} {107}},\ \bibinfo {pages} {126802}
  (\bibinfo {year} {2011})}\BibitemShut {NoStop}%
\bibitem [{\citenamefont {Minnai}\ \emph
  {et~al.}(2017{\natexlab{b}})\citenamefont {Minnai}, \citenamefont
  {Bellacicca}, \citenamefont {Brown},\ and\ \citenamefont
  {Milani}}]{minnai2017facile}%
  \BibitemOpen
  \bibfield  {author} {\bibinfo {author} {\bibfnamefont {C.}~\bibnamefont
  {Minnai}}, \bibinfo {author} {\bibfnamefont {A.}~\bibnamefont {Bellacicca}},
  \bibinfo {author} {\bibfnamefont {S.~A.}\ \bibnamefont {Brown}},\ and\
  \bibinfo {author} {\bibfnamefont {P.}~\bibnamefont {Milani}},\ }\bibfield
  {title} {\bibinfo {title} {Facile fabrication of complex networks of
  memristive devices},\ }\href@noop {} {\bibfield  {journal} {\bibinfo
  {journal} {Scientific reports}\ }\textbf {\bibinfo {volume} {7}},\ \bibinfo
  {pages} {1} (\bibinfo {year} {2017}{\natexlab{b}})}\BibitemShut {NoStop}%
\bibitem [{\citenamefont {Plimpton}(1993)}]{plimpton1993fast}%
  \BibitemOpen
  \bibfield  {author} {\bibinfo {author} {\bibfnamefont {S.}~\bibnamefont
  {Plimpton}},\ }\href@noop {} {\emph {\bibinfo {title} {Fast parallel
  algorithms for short-range molecular dynamics}}},\ \bibinfo {type} {Tech.
  Rep.}\ (\bibinfo  {institution} {Sandia National Labs., Albuquerque, NM
  (United States)},\ \bibinfo {year} {1993})\BibitemShut {NoStop}%
\bibitem [{\citenamefont {Heinz}\ \emph {et~al.}(2008)\citenamefont {Heinz},
  \citenamefont {Vaia}, \citenamefont {Farmer},\ and\ \citenamefont
  {Naik}}]{heinz2008accurate}%
  \BibitemOpen
  \bibfield  {author} {\bibinfo {author} {\bibfnamefont {H.}~\bibnamefont
  {Heinz}}, \bibinfo {author} {\bibfnamefont {R.}~\bibnamefont {Vaia}},
  \bibinfo {author} {\bibfnamefont {B.}~\bibnamefont {Farmer}},\ and\ \bibinfo
  {author} {\bibfnamefont {R.}~\bibnamefont {Naik}},\ }\bibfield  {title}
  {\bibinfo {title} {Accurate simulation of surfaces and interfaces of
  face-centered cubic metals using 12- 6 and 9- 6 lennard-jones potentials},\
  }\href@noop {} {\bibfield  {journal} {\bibinfo  {journal} {The Journal of
  Physical Chemistry C}\ }\textbf {\bibinfo {volume} {112}},\ \bibinfo {pages}
  {17281} (\bibinfo {year} {2008})}\BibitemShut {NoStop}%
\bibitem [{\citenamefont {Stukowski}(2009)}]{stukowski2009visualization}%
  \BibitemOpen
  \bibfield  {author} {\bibinfo {author} {\bibfnamefont {A.}~\bibnamefont
  {Stukowski}},\ }\bibfield  {title} {\bibinfo {title} {Visualization and
  analysis of atomistic simulation data with ovito--the open visualization
  tool},\ }\href@noop {} {\bibfield  {journal} {\bibinfo  {journal} {Modelling
  and Simulation in Materials Science and Engineering}\ }\textbf {\bibinfo
  {volume} {18}},\ \bibinfo {pages} {015012} (\bibinfo {year}
  {2009})}\BibitemShut {NoStop}%
\bibitem [{\citenamefont {Kawamura}\ and\ \citenamefont
  {Leburton}(1993)}]{kawamura1993quantum}%
  \BibitemOpen
  \bibfield  {author} {\bibinfo {author} {\bibfnamefont {T.}~\bibnamefont
  {Kawamura}}\ and\ \bibinfo {author} {\bibfnamefont {J.~P.}~\bibnamefont
  {Leburton}},\ }\bibfield  {title} {\bibinfo {title} {Quantum ballistic
  transport through a double-bend waveguide structure: Effects of disorder},\
  }\href@noop {} {\bibfield  {journal} {\bibinfo  {journal} {Physical Review
  B}\ }\textbf {\bibinfo {volume} {48}},\ \bibinfo {pages} {8857} (\bibinfo
  {year} {1993})}\BibitemShut {NoStop}%
\bibitem [{\citenamefont {Dreher}\ \emph {et~al.}(2005)\citenamefont {Dreher},
  \citenamefont {Pauly}, \citenamefont {Heurich}, \citenamefont {Cuevas},
  \citenamefont {Scheer},\ and\ \citenamefont {Nielaba}}]{dreher2005structure}%
  \BibitemOpen
  \bibfield  {author} {\bibinfo {author} {\bibfnamefont {M.}~\bibnamefont
  {Dreher}}, \bibinfo {author} {\bibfnamefont {F.}~\bibnamefont {Pauly}},
  \bibinfo {author} {\bibfnamefont {J.}~\bibnamefont {Heurich}}, \bibinfo
  {author} {\bibfnamefont {J.~C.}\ \bibnamefont {Cuevas}}, \bibinfo {author}
  {\bibfnamefont {E.}~\bibnamefont {Scheer}},\ and\ \bibinfo {author}
  {\bibfnamefont {P.}~\bibnamefont {Nielaba}},\ }\bibfield  {title} {\bibinfo
  {title} {Structure and conductance histogram of atomic-sized au contacts},\
  }\href@noop {} {\bibfield  {journal} {\bibinfo  {journal} {Physical Review
  B}\ }\textbf {\bibinfo {volume} {72}},\ \bibinfo {pages} {075435} (\bibinfo
  {year} {2005})}\BibitemShut {NoStop}%
\bibitem [{\citenamefont {Kurui}\ \emph {et~al.}(2009)\citenamefont {Kurui},
  \citenamefont {Oshima}, \citenamefont {Okamoto},\ and\ \citenamefont
  {Takayanagi}}]{kurui2009conductance}%
  \BibitemOpen
  \bibfield  {author} {\bibinfo {author} {\bibfnamefont {Y.}~\bibnamefont
  {Kurui}}, \bibinfo {author} {\bibfnamefont {Y.}~\bibnamefont {Oshima}},
  \bibinfo {author} {\bibfnamefont {M.}~\bibnamefont {Okamoto}},\ and\ \bibinfo
  {author} {\bibfnamefont {K.}~\bibnamefont {Takayanagi}},\ }\bibfield  {title}
  {\bibinfo {title} {Conductance quantization and dequantization in gold
  nanowires due to multiple reflection at the interface},\ }\href@noop {}
  {\bibfield  {journal} {\bibinfo  {journal} {Physical Review B}\ }\textbf
  {\bibinfo {volume} {79}},\ \bibinfo {pages} {165414} (\bibinfo {year}
  {2009})}\BibitemShut {NoStop}%
\bibitem [{\citenamefont {Yanson}\ \emph {et~al.}(2005)\citenamefont {Yanson},
  \citenamefont {Shklyarevskii}, \citenamefont {Csonka}, \citenamefont
  {Van~Kempen}, \citenamefont {Speller}, \citenamefont {Yanson},\ and\
  \citenamefont {Van~Ruitenbeek}}]{yanson2005atomic}%
  \BibitemOpen
  \bibfield  {author} {\bibinfo {author} {\bibfnamefont {I.~K.}~\bibnamefont
  {Yanson}}, \bibinfo {author} {\bibfnamefont {O.~I.}~\bibnamefont
  {Shklyarevskii}}, \bibinfo {author} {\bibfnamefont {S.}~\bibnamefont
  {Csonka}}, \bibinfo {author} {\bibfnamefont {H.}~\bibnamefont {van~Kempen}},
  \bibinfo {author} {\bibfnamefont {S.}~\bibnamefont {Speller}}, \bibinfo
  {author} {\bibfnamefont {A.~I.}~\bibnamefont {Yanson}},\ and\ \bibinfo {author}
  {\bibfnamefont {J.~M.}~\bibnamefont {van~Ruitenbeek}},\ }\bibfield  {title}
  {\bibinfo {title} {Atomic-size oscillations in conductance histograms for
  gold nanowires and the influence of work hardening},\ }\href@noop {}
  {\bibfield  {journal} {\bibinfo  {journal} {Physical review letters}\
  }\textbf {\bibinfo {volume} {95}},\ \bibinfo {pages} {256806} (\bibinfo
  {year} {2005})}\BibitemShut {NoStop}%
\bibitem [{\citenamefont {Oshima}\ \emph {et~al.}(2003)\citenamefont {Oshima},
  \citenamefont {Mouri}, \citenamefont {Hirayama},\ and\ \citenamefont
  {Takayanagi}}]{oshima2003development}%
  \BibitemOpen
  \bibfield  {author} {\bibinfo {author} {\bibfnamefont {Y.}~\bibnamefont
  {Oshima}}, \bibinfo {author} {\bibfnamefont {K.}~\bibnamefont {Mouri}},
  \bibinfo {author} {\bibfnamefont {H.}~\bibnamefont {Hirayama}},\ and\
  \bibinfo {author} {\bibfnamefont {K.}~\bibnamefont {Takayanagi}},\ }\bibfield
   {title} {\bibinfo {title} {Development of a miniature stm holder for study
  of electronic conductance of metal nanowires in uhv--tem},\ }\href@noop {}
  {\bibfield  {journal} {\bibinfo  {journal} {Surface science}\ }\textbf
  {\bibinfo {volume} {531}},\ \bibinfo {pages} {209} (\bibinfo {year}
  {2003})}\BibitemShut {NoStop}%
\bibitem [{\citenamefont {Rodrigues}\ \emph {et~al.}(2000)\citenamefont
  {Rodrigues}, \citenamefont {Fuhrer},\ and\ \citenamefont
  {Ugarte}}]{rodrigues2000signature}%
  \BibitemOpen
  \bibfield  {author} {\bibinfo {author} {\bibfnamefont {V.}~\bibnamefont
  {Rodrigues}}, \bibinfo {author} {\bibfnamefont {T.}~\bibnamefont {Fuhrer}},\
  and\ \bibinfo {author} {\bibfnamefont {D.}~\bibnamefont {Ugarte}},\
  }\bibfield  {title} {\bibinfo {title} {Signature of atomic structure in the
  quantum conductance of gold nanowires},\ }\href@noop {} {\bibfield  {journal}
  {\bibinfo  {journal} {Physical review letters}\ }\textbf {\bibinfo {volume}
  {85}},\ \bibinfo {pages} {4124} (\bibinfo {year} {2000})}\BibitemShut
  {NoStop}%
\bibitem [{\citenamefont {Kiguchi}\ \emph {et~al.}(2006)\citenamefont
  {Kiguchi}, \citenamefont {Konishi},\ and\ \citenamefont
  {Murakoshi}}]{kiguchi2006conductance}%
  \BibitemOpen
  \bibfield  {author} {\bibinfo {author} {\bibfnamefont {M.}~\bibnamefont
  {Kiguchi}}, \bibinfo {author} {\bibfnamefont {T.}~\bibnamefont {Konishi}},\
  and\ \bibinfo {author} {\bibfnamefont {K.}~\bibnamefont {Murakoshi}},\
  }\bibfield  {title} {\bibinfo {title} {Conductance bistability of gold
  nanowires at room temperature},\ }\href@noop {} {\bibfield  {journal}
  {\bibinfo  {journal} {Physical Review B}\ }\textbf {\bibinfo {volume} {73}},\
  \bibinfo {pages} {125406} (\bibinfo {year} {2006})}\BibitemShut {NoStop}%
\bibitem [{\citenamefont {Suzuki}\ \emph {et~al.}(2007)\citenamefont {Suzuki},
  \citenamefont {Tsutsui}, \citenamefont {Miura}, \citenamefont {Kurokawa},\
  and\ \citenamefont {Sakai}}]{suzuki2007distribution}%
  \BibitemOpen
  \bibfield  {author} {\bibinfo {author} {\bibfnamefont {R.}~\bibnamefont
  {Suzuki}}, \bibinfo {author} {\bibfnamefont {M.}~\bibnamefont {Tsutsui}},
  \bibinfo {author} {\bibfnamefont {D.}~\bibnamefont {Miura}}, \bibinfo
  {author} {\bibfnamefont {S.}~\bibnamefont {Kurokawa}},\ and\ \bibinfo
  {author} {\bibfnamefont {A.}~\bibnamefont {Sakai}},\ }\bibfield  {title}
  {\bibinfo {title} {Distribution of 1g0 plateau length of au contacts at room
  temperature},\ }\href@noop {} {\bibfield  {journal} {\bibinfo  {journal}
  {Japanese journal of applied physics}\ }\textbf {\bibinfo {volume} {46}},\
  \bibinfo {pages} {3694} (\bibinfo {year} {2007})}\BibitemShut {NoStop}%
\bibitem [{\citenamefont {Kizuka}(2008)}]{kizuka2008atomic}%
  \BibitemOpen
  \bibfield  {author} {\bibinfo {author} {\bibfnamefont {T.}~\bibnamefont
  {Kizuka}},\ }\bibfield  {title} {\bibinfo {title} {Atomic configuration and
  mechanical and electrical properties of stable gold wires of single-atom
  width},\ }\href@noop {} {\bibfield  {journal} {\bibinfo  {journal} {Physical
  review B}\ }\textbf {\bibinfo {volume} {77}},\ \bibinfo {pages} {155401}
  (\bibinfo {year} {2008})}\BibitemShut {NoStop}%
\bibitem [{\citenamefont {Yasuda}\ and\ \citenamefont
  {Sakai}(1997)}]{yasuda1997conductance}%
  \BibitemOpen
  \bibfield  {author} {\bibinfo {author} {\bibfnamefont {H.}~\bibnamefont
  {Yasuda}}\ and\ \bibinfo {author} {\bibfnamefont {A.}~\bibnamefont {Sakai}},\
  }\bibfield  {title} {\bibinfo {title} {Conductance of atomic-scale gold
  contacts under high-bias voltages},\ }\href@noop {} {\bibfield  {journal}
  {\bibinfo  {journal} {Physical Review B}\ }\textbf {\bibinfo {volume} {56}},\
  \bibinfo {pages} {1069} (\bibinfo {year} {1997})}\BibitemShut {NoStop}%
\bibitem [{\citenamefont {Soler}\ \emph {et~al.}(2002)\citenamefont {Soler},
  \citenamefont {Artacho}, \citenamefont {Gale}, \citenamefont {Garc{\'\i}a},
  \citenamefont {Junquera}, \citenamefont {Ordej{\'o}n},\ and\ \citenamefont
  {S{\'a}nchez-Portal}}]{soler2002siesta}%
  \BibitemOpen
  \bibfield  {author} {\bibinfo {author} {\bibfnamefont {J.~M.}\ \bibnamefont
  {Soler}}, \bibinfo {author} {\bibfnamefont {E.}~\bibnamefont {Artacho}},
  \bibinfo {author} {\bibfnamefont {J.~D.}\ \bibnamefont {Gale}}, \bibinfo
  {author} {\bibfnamefont {A.}~\bibnamefont {Garc{\'\i}a}}, \bibinfo {author}
  {\bibfnamefont {J.}~\bibnamefont {Junquera}}, \bibinfo {author}
  {\bibfnamefont {P.}~\bibnamefont {Ordej{\'o}n}},\ and\ \bibinfo {author}
  {\bibfnamefont {D.}~\bibnamefont {S{\'a}nchez-Portal}},\ }\bibfield  {title}
  {\bibinfo {title} {The siesta method for ab initio order-n materials
  simulation},\ }\href@noop {} {\bibfield  {journal} {\bibinfo  {journal}
  {Journal of Physics: Condensed Matter}\ }\textbf {\bibinfo {volume} {14}},\
  \bibinfo {pages} {2745} (\bibinfo {year} {2002})}\BibitemShut {NoStop}%
\bibitem [{\citenamefont {Brandbyge}\ \emph {et~al.}(2002)\citenamefont
  {Brandbyge}, \citenamefont {Mozos}, \citenamefont {Ordej{\'o}n},
  \citenamefont {Taylor},\ and\ \citenamefont
  {Stokbro}}]{brandbyge2002density}%
  \BibitemOpen
  \bibfield  {author} {\bibinfo {author} {\bibfnamefont {M.}~\bibnamefont
  {Brandbyge}}, \bibinfo {author} {\bibfnamefont {J.-L.}\ \bibnamefont
  {Mozos}}, \bibinfo {author} {\bibfnamefont {P.}~\bibnamefont {Ordej{\'o}n}},
  \bibinfo {author} {\bibfnamefont {J.}~\bibnamefont {Taylor}},\ and\ \bibinfo
  {author} {\bibfnamefont {K.}~\bibnamefont {Stokbro}},\ }\bibfield  {title}
  {\bibinfo {title} {Density-functional method for nonequilibrium electron
  transport},\ }\href@noop {} {\bibfield  {journal} {\bibinfo  {journal}
  {Physical Review B}\ }\textbf {\bibinfo {volume} {65}},\ \bibinfo {pages}
  {165401} (\bibinfo {year} {2002})}\BibitemShut {NoStop}%
\bibitem [{\citenamefont {Landauer}(1978)}]{landauer1978}%
  \BibitemOpen
  \bibfield  {author} {\bibinfo {author} {\bibfnamefont {R.}~\bibnamefont
  {Landauer}},\ }\bibfield  {title} {\bibinfo {title} {Electrical conductivity
  in inhomogeneous media},\ }\href {https://doi.org/10.1063/1.31150} {\bibfield
   {journal} {\bibinfo  {journal} {AIP Conference Proceedings}\ }\textbf
  {\bibinfo {volume} {40}},\ \bibinfo {pages} {2} (\bibinfo {year}
  {1978})}\BibitemShut {NoStop}%
\bibitem [{\citenamefont {Erts}\ \emph {et~al.}(2000)\citenamefont {Erts},
  \citenamefont {Olin}, \citenamefont {Ryen}, \citenamefont {Olsson},\ and\
  \citenamefont {Th{\"o}l{\'e}n}}]{erts2000maxwell}%
  \BibitemOpen
  \bibfield  {author} {\bibinfo {author} {\bibfnamefont {D.}~\bibnamefont
  {Erts}}, \bibinfo {author} {\bibfnamefont {H.}~\bibnamefont {Olin}}, \bibinfo
  {author} {\bibfnamefont {L.}~\bibnamefont {Ryen}}, \bibinfo {author}
  {\bibfnamefont {E.}~\bibnamefont {Olsson}},\ and\ \bibinfo {author}
  {\bibfnamefont {A.}~\bibnamefont {Th{\"o}l{\'e}n}},\ }\bibfield  {title}
  {\bibinfo {title} {Maxwell and sharvin conductance in gold point contacts
  investigated using tem-stm},\ }\href@noop {} {\bibfield  {journal} {\bibinfo
  {journal} {Physical Review B}\ }\textbf {\bibinfo {volume} {61}},\ \bibinfo
  {pages} {12725} (\bibinfo {year} {2000})}\BibitemShut {NoStop}%
\bibitem [{\citenamefont {Stauffer}\ and\ \citenamefont
  {Aharony}(1994)}]{stauffer1994}%
  \BibitemOpen
  \bibfield  {author} {\bibinfo {author} {\bibfnamefont {D.}~\bibnamefont
  {Stauffer}}\ and\ \bibinfo {author} {\bibfnamefont {A.}~\bibnamefont
  {Aharony}},\ }\href {https://books.google.it/books?id=v66plleij5QC} {\emph
  {\bibinfo {title} {Introduction To Percolation Theory}}}\ (\bibinfo
  {publisher} {Taylor \& Francis},\ \bibinfo {year} {1994})\BibitemShut
  {NoStop}%
\bibitem [{\citenamefont {Maaroof}\ and\ \citenamefont
  {Evans}(1994)}]{maaroof1994onset}%
  \BibitemOpen
  \bibfield  {author} {\bibinfo {author} {\bibfnamefont {A.}~\bibnamefont
  {Maaroof}}\ and\ \bibinfo {author} {\bibfnamefont {B.}~\bibnamefont
  {Evans}},\ }\bibfield  {title} {\bibinfo {title} {Onset of electrical
  conduction in pt and ni films},\ }\href@noop {} {\bibfield  {journal}
  {\bibinfo  {journal} {Journal of applied Physics}\ }\textbf {\bibinfo
  {volume} {76}},\ \bibinfo {pages} {1047} (\bibinfo {year}
  {1994})}\BibitemShut {NoStop}%
\bibitem [{\citenamefont {Burgmann}\ \emph {et~al.}(2005)\citenamefont
  {Burgmann}, \citenamefont {Lim}, \citenamefont {McCulloch}, \citenamefont
  {Gan}, \citenamefont {Davies}, \citenamefont {McKenzie},\ and\ \citenamefont
  {Bilek}}]{burgmann2005electrical}%
  \BibitemOpen
  \bibfield  {author} {\bibinfo {author} {\bibfnamefont {F.}~\bibnamefont
  {Burgmann}}, \bibinfo {author} {\bibfnamefont {S.}~\bibnamefont {Lim}},
  \bibinfo {author} {\bibfnamefont {D.}~\bibnamefont {McCulloch}}, \bibinfo
  {author} {\bibfnamefont {B.}~\bibnamefont {Gan}}, \bibinfo {author}
  {\bibfnamefont {K.}~\bibnamefont {Davies}}, \bibinfo {author} {\bibfnamefont
  {D.~R.}\ \bibnamefont {McKenzie}},\ and\ \bibinfo {author} {\bibfnamefont
  {M.}~\bibnamefont {Bilek}},\ }\bibfield  {title} {\bibinfo {title}
  {Electrical conductivity as a measure of the continuity of titanium and
  vanadium thin films},\ }\href@noop {} {\bibfield  {journal} {\bibinfo
  {journal} {Thin Solid Films}\ }\textbf {\bibinfo {volume} {474}},\ \bibinfo
  {pages} {341} (\bibinfo {year} {2005})}\BibitemShut {NoStop}%
\bibitem [{\citenamefont {Barborini}\ \emph {et~al.}(2010)\citenamefont
  {Barborini}, \citenamefont {Corbelli}, \citenamefont {Bertolini},
  \citenamefont {Repetto}, \citenamefont {Leccardi}, \citenamefont {Vinati},\
  and\ \citenamefont {Milani}}]{barborini2010influence}%
  \BibitemOpen
  \bibfield  {author} {\bibinfo {author} {\bibfnamefont {E.}~\bibnamefont
  {Barborini}}, \bibinfo {author} {\bibfnamefont {G.}~\bibnamefont {Corbelli}},
  \bibinfo {author} {\bibfnamefont {G.}~\bibnamefont {Bertolini}}, \bibinfo
  {author} {\bibfnamefont {P.}~\bibnamefont {Repetto}}, \bibinfo {author}
  {\bibfnamefont {M.}~\bibnamefont {Leccardi}}, \bibinfo {author}
  {\bibfnamefont {S.}~\bibnamefont {Vinati}},\ and\ \bibinfo {author}
  {\bibfnamefont {P.}~\bibnamefont {Milani}},\ }\bibfield  {title} {\bibinfo
  {title} {The influence of nanoscale morphology on the resistivity of
  cluster-assembled nanostructured metallic thin films},\ }\href@noop {}
  {\bibfield  {journal} {\bibinfo  {journal} {New Journal of Physics}\ }\textbf
  {\bibinfo {volume} {12}},\ \bibinfo {pages} {073001} (\bibinfo {year}
  {2010})}\BibitemShut {NoStop}%
\bibitem [{\citenamefont {Namba}(1970)}]{Namba_1970}%
  \BibitemOpen
  \bibfield  {author} {\bibinfo {author} {\bibfnamefont {Y.}~\bibnamefont
  {Namba}},\ }\bibfield  {title} {\bibinfo {title} {Resistivity and temperature
  coefficient of thin metal films with rough surface},\ }\href
  {https://doi.org/10.1143/jjap.9.1326} {\bibfield  {journal} {\bibinfo
  {journal} {Japanese Journal of Applied Physics}\ }\textbf {\bibinfo {volume}
  {9}},\ \bibinfo {pages} {1326} (\bibinfo {year} {1970})}\BibitemShut
  {NoStop}%
\bibitem [{\citenamefont {Munoz}\ and\ \citenamefont
  {Arenas}(2017)}]{munoz2017size}%
  \BibitemOpen
  \bibfield  {author} {\bibinfo {author} {\bibfnamefont {R.~C.}\ \bibnamefont
  {Munoz}}\ and\ \bibinfo {author} {\bibfnamefont {C.}~\bibnamefont {Arenas}},\
  }\bibfield  {title} {\bibinfo {title} {Size effects and charge transport in
  metals: Quantum theory of the resistivity of nanometric metallic structures
  arising from electron scattering by grain boundaries and by rough surfaces},\
  }\href@noop {} {\bibfield  {journal} {\bibinfo  {journal} {Applied Physics
  Reviews}\ }\textbf {\bibinfo {volume} {4}},\ \bibinfo {pages} {011102}
  (\bibinfo {year} {2017})}\BibitemShut {NoStop}%
\bibitem [{\citenamefont {Beloborodov}\ \emph {et~al.}(2007)\citenamefont
  {Beloborodov}, \citenamefont {Lopatin}, \citenamefont {Vinokur},\ and\
  \citenamefont {Efetov}}]{beloborodov2007}%
  \BibitemOpen
  \bibfield  {author} {\bibinfo {author} {\bibfnamefont {I.~S.}\ \bibnamefont
  {Beloborodov}}, \bibinfo {author} {\bibfnamefont {A.~V.}\ \bibnamefont
  {Lopatin}}, \bibinfo {author} {\bibfnamefont {V.~M.}\ \bibnamefont
  {Vinokur}}, \ and\ \bibinfo {author} {\bibfnamefont {K.~B.}\ \bibnamefont
  {Efetov}},\ }\href {\doibase 10.1103/RevModPhys.79.469} {\bibfield  {journal}
  {\bibinfo  {journal} {Rev. Mod. Phys.}\ }\textbf {\bibinfo {volume} {79}},\
  \bibinfo {pages} {469} (\bibinfo {year} {2007})}\BibitemShut {NoStop}%
\end{thebibliography}

%

\end{document}


\title{Supplemental material to the manuscript  MG10236}
\maketitle

\section{Tunneling and single grain conductance}

In modeling electron transport in granular materials,
one can distinguish different transport regimes
using two key quantities \cite{beloborodov2007}:
the intergranular conductance $g$ and the intragrain conductance $g_0$,
both expressed in units of the quantum conductance $e^2/\hbar$.
In the following we will provide an estimate of those quantities
and prove that our system is in a conducting ( $g_0 \gg 1$) and strong grain-coupling
($g\gg 1$) regime, in which effects such as Coulomb blockade can be excluded.

The intragrain conductance can be estimated
as $g_0=E_{th}/\delta$ \cite{beloborodov2007}, 
where $E_{th}$ is the Thouless energy and $\delta$ 
the mean energy level spacing for a single grain.
Under the simplifying assumption that our system
is made of Au clusters of diameter  $d=8.8\; nm$
(instead of two distinct populations with  $d=8.8\; nm$ and  $d=1.3 \;nm$),
the mean energy level spacing can be estimated to be $\delta\sim 20\;meV$ \cite{barmparis2016},
while $E_{th}=l_e v_F/d= 800\; meV$, 
where we have considered an electron mean free path $l_e= 40\; nm$ 
and a Fermi velocity $v_F 1.4\cdot 106 \;m/s$, both at $T=300K$.
Combining those values one obtains $g_0=40$, which indeed is much greater than the unity
and proves that are Au clusters are indeed good conductors.

The intergranular (also called ``tunneling'', 
even when the regime is not of \emph{quantum} tunneling) conductance $g$
is harder to estimate from the available information on the experiment only.
We therefore resort to our simulation and ab initio calculations to provide
an estimate of the conductance of junctions as the one represented in the following picture:
\begin{figure}[h!]
  \includegraphics[width=5cm]{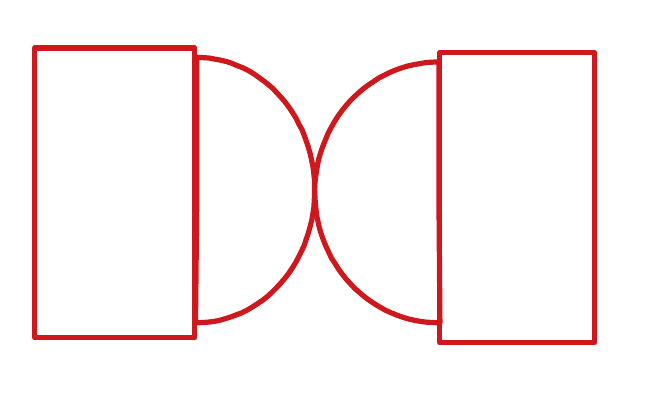}
\end{figure}\\
where the outer rectangles identify the electrodes and 
the inner hemispheres the interface between two nanoparticles. 
In a granular sample composed by mono-sized deformable but incompressible 
(i.e. volume conserving) spheres,
the average surface of contact between two grains $S_{avg}$
is only function of the occupied volume fraction $\Phi$.
Considering $\Phi=0.675$, which is extracted from our simulated sample,
we can estimate $S_{avg}=15\; nm^2$, based on systems with similar geometrical characteristics
\cite{PhysRevE.84.011302}. Given $S_{avg}$ the conductance of the corresponding junction
can be extrapolated from our ab initio calculations for similar junctions.
In particular, using the data for non-fcc interfaces shown in Figure 4 of the manuscript,
we compute a tunneling conductance of $g\sim 30$, 
from which we can deduce that our system is in the strong coupling regime.
As the ratio between grain and tunneling conductance is $g_0/g\sim1.3$, 
we can say that our system is so strongly coupled that can be considered as a porous metal.

\section{Electron Localization}

The inverse of the Thouless energy $1/E_{th}=1.25 \;eV^{-1}$ gives an estimate of 
the time that it takes for an electron to traverse the grain.
By comparing that with the time takes an electron to pass from a grain to the next one,
which can be estimated as $(g\delta)^{-1}=1.7 \;eV^{-1}$ \cite{beloborodov2007},
we can say that electrons are effectively delocalized
(they time they spend in a grain is roughly the time
they need to go from one edge to the other)
and effects such as Coulomb blockade can be excluded.

\section{Sample conductivity from granular model}

The conductivity of dense nanoparticle systems can be estimated using
a model of ordered arrays of conducting, strongly coupled grains,
which can be solved analytically \cite{strong1,strong2}.
Using the parameters discussed above, 
one obtains $\sigma \approx 5.246\times 10^5 \Omega^{-1} m^{-1}$.
To see how this compares with our estimate, we superimpose the corresponding resistance
to Fig. 6 of the manuscript, as follows:
\begin{figure}[h!]
  \includegraphics[width=13cm]{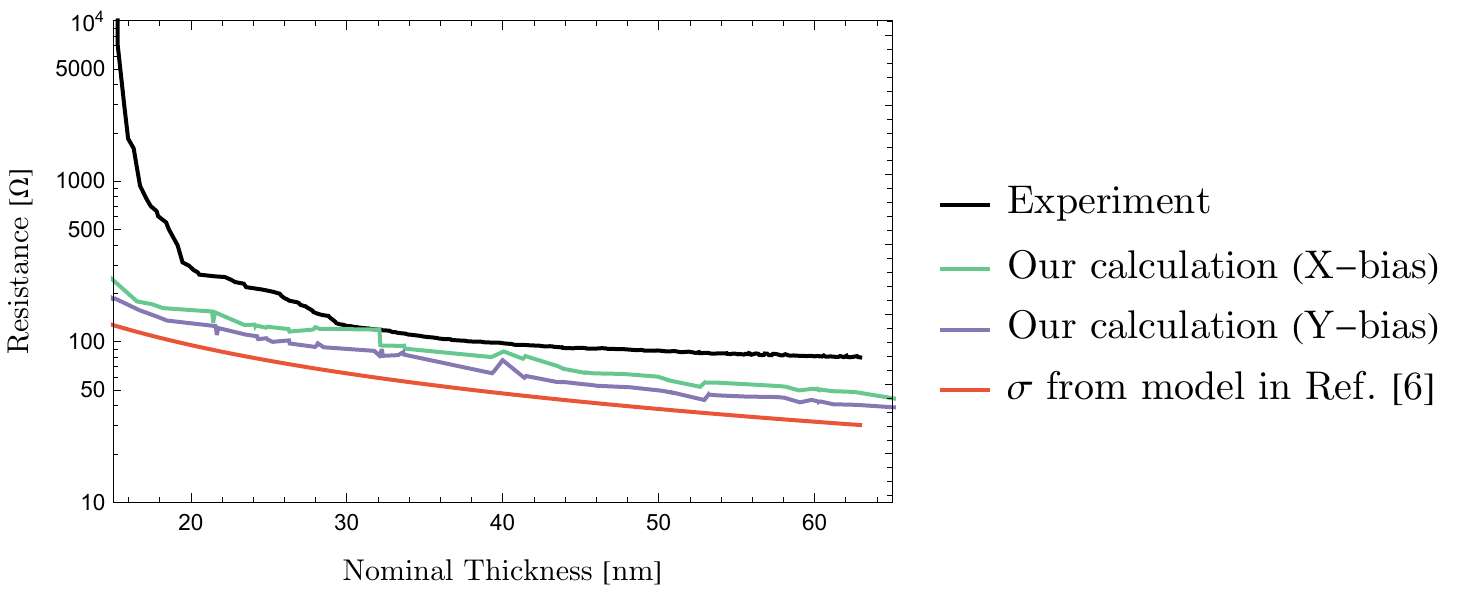}
\end{figure}\\
We observe an underestimation of the resistance that is slightly stronger than our own estimate. 
We would therefore argue that this can be regarded as a further validation of our approach.

The model includes corrections to the classical Drude conductivity for a granular metal
due to Coulomb interaction, which here are found to be negligible. 
More specifically, in writing the conductivity as
$\sigma=\sigma_0+\delta \sigma_1+\delta \sigma_2$, where $\sigma_0$ is the Drude conductivity,
$\delta \sigma_1$ and $\delta\sigma_2$
the correction due to the contribution from the large and low energy scales, respectively,
we found that $\sigma_0=5.25\times 10^5 \Omega^{-1} m^{-1}$,
$\delta \sigma_1=-474.2 \;\Omega^{-1}m^{-1}$ and $\delta \sigma_2=56.3\;\Omega^{-1}m^{-1}$,
which shows that these corrections are both smaller than $\sigma_0$ by orders of magnitude.
We can therefore deduce that Coulomb interaction and quantum interference can be neglected.
Finally, we notice that the Coulomb energy (needed to compute $\sigma_1$) $E_c=e^2/(2C)$, 
with $C=4\pi \epsilon_0 \epsilon_r d$ and $\epsilon_r =6.9$ for Au, 
gives an estimate of $E_c\sim 12\; meV$, which is lower than the thermal energy $26 \;meV$.
